\documentclass[preprint,showpacs,preprintnumbers,amsmath,amssymb,nofootinbib]{revtex4}

\usepackage{etex}
\usepackage{amsthm,amscd,amsbsy,array}
\usepackage{bm}
\usepackage{soul} 

\usepackage{graphics,graphicx,xcolor}

\usepackage[colorlinks=true, pdfstartview=FitV, linkcolor=blue, citecolor=blue, urlcolor=blue]{hyperref} 

\newcommand{\red}{\textcolor{red}}
\newcommand{\blue}{\textcolor{blue}}

\newcommand{\gb}{\colorbox{green}}

\newenvironment{redtext}{\color{red}}{\ignorespacesafterend}
\newenvironment{bluetext}{\color{blue}}{\ignorespacesafterend}
\newenvironment{greentext}{\color{green}}{\ignorespacesafterend}
\newenvironment{magentatext}{\color{magenta}}{\ignorespacesafterend}

\newcommand{\bmag}{\begin{magentatext}}
\newcommand{\emag}{\end{magentatext}}
\newcommand{\bcyan}{\begin{cyantext}}
\newcommand{\ecyan}{\end{cyantext}}

\newcommand{\bblue}{\begin{bluetext}}
\newcommand{\eblue}{\end{bluetext}}
\newcommand{\bred}{\begin{redtext}}
\newcommand{\ered}{\end{redtext}}
\newcommand{\bgreen}{\begin{greentext}}
\newcommand{\egreen}{\end{greentext}}
\newcommand{\bdgreen}{\begin{dgreentext}}
\newcommand{\edgreen}{\end{dgreentext}}
\numberwithin{equation}{section}

\let\ssection=\section
\renewcommand{\section}{\setcounter{equation}{0}\ssection}

\newcommand{\ba}{{\bf a}}

\newcommand{\cJ}{{\mathcal{J}}}

\newcommand{{\cL}}{{\mathcal{L}}}

\newcommand{\cQ}{{Q}}

\newcommand{\bx}{{\bm{x}}}

\def\vnabla{{\bm{\nabla}}}

\def\smallover#1/#2{\hbox{$\textstyle\frac{#1}{#2}$}} %

\def\where{{\quad\text{where}\quad}}

\def\benu{\begin{enumerate}}
\def\eenu{\end{enumerate}}
\def\beq{\begin{equation}}
\def\eeq{\end{equation}}
\def\beqa{\begin{eqnarray}}
\def\eeqa{\end{eqnarray}}

\def\barray{\left(\begin{array}}
\def\earray{\end{array}\right)}
\def\barraynb{\begin{array}}
\def\earraynb{\end{array}}

\def\?{\quad{\gb{\fbox{\texttt{?}}\;}}\quad}
\def\p{{\partial}}

\def\v0{\mathbf{0}}

\def\Rarrow{{\quad\Rightarrow\quad}}

\newcommand{\mH}{{\mathcal H}}
\newcommand{\mK}{{\mathcal K}}
\newcommand{\mL}{{\mathcal L}}
\newcommand{\mA}{{\mathcal A}}

\def\G11{\Gamma_{11} }

\newcommand{\const}{\mathop{\rm const.}\nolimits}
\newcommand{\half }{\frac{1}{2}}

\def\implies{\Rightarrow}

\def\smallover#1/#2{\hbox{$\textstyle\frac{#1}{#2}$}} %
\def\smallcirc{{\raise 0.5pt \hbox{$\scriptstyle\circ$}}}
\def\2{{\smallover1/2}}

\newcommand{\bigbox}[1]{\fbox{%
\rule[-20pt]{0pt}{45pt}$\;\;\displaystyle{#1}\;\;$}
}
\newcommand{\medbox}[1]{\fbox{%
\rule[-10pt]{0pt}{25pt}$\;\;\displaystyle{#1}\;\;$}%
}

\let\ssection=\section
\renewcommand{\section}
{\setcounter{equation}{0}\ssection}

\def\besub{\begin{subequations}}
\def\esub{\end{subequations}}
\begin{document}

\preprint{\texttt{arXiv:2003.07649}}

\title{Conformal symmetries and integrals of the motion in pp waves with external electromagnetic fields}

\author{
K. Andrzejewski$^{1}$\footnote{email: krzysztof.andrzejewski@uni.lodz.pl},
N. Dimakis${}^{2}$\footnote{e-mail:nsdimakis@gmail.com},
M. Elbistan${}^{3,4}$\footnote{mailto:mahmut.Elbistan@lmpt.univ-tours.fr},
P. A. Horvathy${}^{3}$\footnote{mailto:horvathy@lmpt.univ-tours.fr},
P. Kosi\'nski$^{1}$\footnote{email: pkosinsk@uni.lodz.pl},
P.-M. Zhang${}^{4}$\footnote{
e-mail:zhangpm5@mail.sysu.edu.cn
}
}
\affiliation{
${}^1$
Department of Computer Science, Faculty of Physics and Applied Informatics\\
University of L\'od\'z,
Pomorska 149/153, 90-236 L\'od\'z, Poland
\\
${}^{2}$ Center for Theoretical Physics, College of Physical Science and Technology, Sichuan University, Chengdu 610065, China
\\
${}^3$  Institut Denis Poisson CNRS/UMR 7013 - Universit\'e de Tours - Universit\'e d'Orl\'eans Parc de Grandmont, 37200, Tours, (France)
\\
${}^4$ School of Physics and Astronomy, Sun Yat-sen University, Zhuhai, China
}

\date{\today}

\pacs{\\
11.30.-j  Symmetry and conservation laws\\
04.20.-q  Classical general relativity\\
04.30.-w  Gravitational waves
}

\begin{abstract}
The integrals of the motion associated with conformal Killing vectors of a curved space-time with an additional electromagnetic background are studied  for massive particles. They involve a new term which might be  non-local. The difficulty disappears for  pp-waves, for which  explicit, local  conserved charges are found.  
Alternatively, the mass can be taken into account by ``distorting'' the conformal Killing vectors. The relation of these non-point symmetries to the charges is analysed both in the Lagrangian and Hamiltonian approaches, as well as in the framework of Eisenhart-Duval lift.
\\[6pt]

\noindent
{Annals of Physics 418 (2020) 168180. 
https://doi.org/10.1016/j.aop.2020.168180}

\end{abstract}

\maketitle

\tableofcontents

\newpage
\section{Introduction}\label{Intro}

Conserved quantities associated with the {isometries} play an important role for the integrability of the geodesic equations~:
 the Killing vectors  generate Noether symmetries and provide, consequently, integrals of the motion. (The argument can be extended to electromagnetic backgrounds, provided  the latter  are also preserved).  Similar results hold for homothetic fields whose conformal factor is a constant \cite{Henkel,Igata,Maughan,Zhang:2019koe}, or for massless geodesics   \cite{CGHHZ,Harte,Conf4GW}. Recent applications involve  the so called  \emph{Memory Effect} for gravitational waves, see \cite{Memory,Ilderton,OurMemory} and references therein.
However the procedure above breaks down for  \emph{proper conformal Killing vectors} whose conformal factor is not a constant and for massive geodesics.

Recently an alternative, non-Noetherian,  approach to this problem has been  proposed \cite{b2}. It has been shown that conformal Killing fields lead, except for some special parametrisation, to \emph{non-local integrals of the motion}~: the associated charge  involves  a novel type \emph{integral term},
\beq
m\!\int\! \omega_Y(x(\tau))\sqrt{-g_{\mu\nu}\dot x^\mu\dot x^\nu}\,d\tau\,.
\label{intterm}
\eeq
Calculating (\ref{intterm}) requires to solve first the geodesic equations, making its use difficult.

The theory of ref. \cite{b2} is indeed the staring point for our considerations here. First  we extend the result above to  conformal fields $Y$   ($\mL_Y g=2\omega_Y g  $)  which preserve also some electromagnetic background given by a potential $A$, i.e.,
\beq
 \mL_Y A=d\phi_Y, \quad (\mL_YF=0)
\label{e3}
\eeq
for some function  $\phi_Y$. For a geodesic of mass $m$ the resulting charge is given in eqn. (\ref{e4}) below.

For a pp wave the charge can be calculated explicitly, raising several questions. Firstly, what is the relation (if any)  between  the conserved  charges for pp-waves obtained above and the Noetherian   or  non-point-symmetry approach to the geodesic equations~?
Secondly, can  such  charges provide  new information or are they functions of known charges~?
Thirdly, can we find electromagnetic backgrounds preserved by the conformal fields and  the explicit forms of the corresponding integrals of the motion~?

Let us now recall that according to the Eisenhart-Duval approach  \cite{Eisenhart,Bargmann,DGH91,BekaertMorand,dissip}  the  two-dimensional classical  dynamics can be lifted to massless geodesics, allowing us to recover the classical non-relativistic symmetries ``downstairs'' from those, relativistic, ``upstairs''. Can we describe the charges obtained above and find their meaning within this framework~?

This paper is devoted to  answering  these questions.

An important observation is that for massive geodesics proper conformal vector fields $Y^\alpha$ \emph{can not be  realised as a Noether point  symmetries. However we show  that they can be generated instead  by \emph{``distorting''} $Y^\alpha$ in a mass-dependent way,
\beq
\Upsilon^\alpha = Y^\alpha + \frac{m^2}{p_v^2} f^\alpha\,,
\eeq
where $p_v$ is the conserved ``vertical'' momentum and  $f^\alpha$ is suitably defined vector field} cf. \eqref{upsdef} -- so that  
the associated conserved charge reproduces
the ``non-local'' one which will be constructed in sec.\ref{IntConf}, cf. \eqref{e4}. 
In particular, the integral term
\eqref{intterm} is induced by ``distorsion vector field'' $f^{\alpha}$.

These ``distorted symmetries''  are reminiscent of  dynamical symmetries  (whose typical example is  the ${\rm o}(4)$ symmetry of the Kepler problem involving the Laplace-Runge-Lenz vector)  in that
they are \emph{not point symmetries} as (say) rotations or translations. Their generators involve also derivatives of the configuration space variables.

Our paper is organized as follows. In sec.  \ref{IntConf} we generalize the results obtained in \cite{Henkel,b2,Igata} to  proper conformal fields and electromagnetic backgrounds preserved  by them; we present both Lagrangian and Hamiltonian approaches.
The explicit form of integrals of the motion for  pp-waves is spelled out in sec. \ref{Confpp}. The relation to the Eisenhart-Duval lift \cite{Eisenhart,Bargmann,DGH91,BekaertMorand,dissip}  as well as  further properties of the  charges are obtained. Examples  of  electromagnetic backgrounds are presented in sec. \ref{furtherStud}.
The  relations between distorted symmetries and local charges is studied in sec. \ref{ModConfpp}.
Further illustrations  are contained   in sec. \ref{ExamplesSec}.
\goodbreak

\section{``Non-local'' integrals of the motion
}\label{IntConf}
A spinless particle in a relativistic space-time in the presence of additional  vector potential  $A_\mu$ is described by  the Lagrangian
\beq
\label{e1}
L=-m\sqrt{-g_{\mu\nu}\dot x^\mu\dot x^\nu}+eA_\mu\dot x^\mu\,,
\eeq
which implies the equations of  the motion
\beq
\label{e2}
m(\overset{..}{x}^\mu+\Gamma^\mu_{\beta\alpha}\dot x^\beta\dot x^\alpha)=m \frac{\;\; d}{d\tau}\left(\ln\sqrt{- g_{\alpha\beta}\dot x^\alpha\dot x^\beta}\right) \dot x^\mu+e\sqrt{- g_{\alpha\beta}\dot x^\alpha\dot x^\beta}{F^\mu}_\nu\dot x^\nu\,,
\eeq
where the ``dot''  means derivation w.r.t. an arbitrary parameter $\tau$.
\par
In terms of constrained systems \cite{Dirac,Sund}  \eqref{e1} is a singular Lagrangian and, due \emph{parametrisation invariance},
the equations of the motion \eqref{e2} are not all independent from each other~: the  action remains form-invariant under the infinite dimensional group of transformations $\tau \mapsto \widetilde{\tau}=f(\tau)$ where $f$ is an arbitrary function. Then, according to the  \emph{second Noether theorem}  \cite{Sund,Noether}, the equations of the motion satisfy an identity.
Reparametrisation invariance implies that in a $d$-dimensional space-time
only $d-1$ of the  $x^\mu$ are independent;  and additional conditions can be imposed.

Let us now assume that $Y$ is a conformal vector for the metric which leaves up-to-a-gauge transformation invariant also  the vector potential, eqn. \eqref{e3}.
Then a tedious calculation shows that for a conformal transformation $Y$ with conformal factor $\omega\equiv\omega_Y$ and $\phi\equiv\phi_Y$, the quantity
\beq
\label{e4}
\bigbox{
I\equiv I_Y=\frac{m}{\sqrt{-g_{\alpha\beta}\dot x^\alpha\dot x^\beta}}Y_\mu\dot x^\mu+m\!\int\! \omega(x(\tau))\sqrt{-g_{\mu\nu}\dot x^\mu\dot x^\nu}\,d\tau+eY_\mu A^\mu-e\phi\,}
\eeq
is a constant of the motion~: $\dot{I}=0$  along each  trajectory, extending  to  electromagnetic fields the  results presented in \cite{b2}. 
For an  {isometry} ($\omega\equiv0$), the non-local term vanishes and the well known result is recovered. Similarly, the (local) charge for a massless particle is obtained in
the (singular) limit $m\to0$.

The main disadvantage  of such an integral of the motion is that it is in general \emph{non local}: the explicit form of the trajectory is needed to calculate it.
The charge may become local using a special parametrisation. We can, for example,  perform the transformation \cite{b2}
$
s=s(\tau)=\displaystyle\int^\tau\!(-g_{\alpha\beta}\dot x^\alpha\dot x^\beta)^{1/2}\,\omega(x(\widetilde{\tau}))d\widetilde{\tau}\,
$ yielding a local expression,
\beq
\label{e5}
\sqrt{-g_{\mu\nu}\frac{d x^\mu}{ds }\frac{d x^\nu}{ds}}=\frac{1}{\omega(x(s))}\,
\Rarrow
I=m(\omega Y_\mu\frac{dx^\mu}{ds}+s)+eY_\mu A^\mu-e\phi\,;
\eeq
the price to pay in this generic time gauge is that the meaning of the geodesic equations becomes obscured.

A  frequent choice is that of an  \emph{affine parameter} $\sigma$, characterized by the property
\beq
g_{\mu\nu}  \frac{dx^\mu}{d\sigma } \frac{d x^\nu}{d\sigma}= -m^2.
 \label{affinecond}
 \eeq
Then the equations of the motion \eqref{e2} resp. the conserved quantity \eqref{e4} become
\besub
\begin{align}
\label{e2aff}
&\frac{d^2 x^\mu}{d\sigma^2}+\Gamma^\mu_{\beta\alpha} \frac{d x^\beta}{d\sigma} \frac{d x^\alpha}{d\sigma}=eF^\mu_{\ \nu}\frac{d x^\nu}{d\sigma}\,,
\\[6pt]
\label{e4aff}
&I=Y_\mu\frac{d x^\mu}{d\sigma}+m^2\!\int^\sigma\!\omega(x(\tilde{\sigma}))d\tilde{\sigma} \,+eY_\mu A^\mu-e\phi\,.
\end{align}
\label{affineqs}
\esub
To conclude this section let us have a look  from the Hamiltonian point of view at the integral of the motion in an arbitrary parametrisation.
The  Lagrangian \eqref{e1} leads to an identically vanishing  Hamiltonian. In order to generate dynamics in the phase space, we  extend  the configuration space by adding a new  coordinate $N$ (the Einbein \cite{Einbein})  and define  an equivalent quadratic Lagrangian
\beq
\label{e8}
\widetilde{L}=\frac{1}{2N}g_{\alpha\beta}\dot x^\alpha\dot x^\beta-\frac{m^2}{2}N \,+ eA_\alpha\dot x^\alpha\,,
\eeq
whose E-L equations reproduce \eqref{e2}.
In terms of the Euler derivatives $E_N$ and $E_\alpha$ for the degrees of freedom $N$ and $x^\alpha$, respectively,
the Euler-Lagrange system consists of $d+1$ equations. One of them, namely $E_N(\widetilde{L})=\frac{\partial \widetilde{L}}{\partial N}=0$, is a constraint. Thus only $d-1$ of the remaining $d$ equations $E_\alpha(\widetilde{L})=0$, are independent~:
the constraint which involves only velocities sets a restriction among the $d$ second order equations, reducing the number of truly independent relations. 

The situation is analogous to what happens for Einstein's equations in four dimensions~: you have 10 equations, 4 of which are constraints. As a consequence, of the 6 remaining equations only 6-4=2 are truly independent~: General Relativity has two physical degrees of freedom.

Both cases are effected by Noether's second theorem and the existence of identities amongst the equations of motion. 

The new variable  $N$ corresponds, through  $E_N(\widetilde{L})=0$, to
\beq
N(\tau)=m^{-1}\sqrt{-g_{\alpha\beta}\dot x^\alpha\dot x^\beta}\,.
\label{Np}
\eeq
Thus, putting $p_\alpha=N^{-1} g_{\alpha\beta}\dot x^\beta+eA_\alpha$,
\beq
\label{e9}
\medbox{
I=Y^\alpha p_\alpha+m^2\!\int\!\!\omega(x(\tau))N(\tau) d\tau\,-e\phi
}
\eeq
is an integral of the motion, as confirmed  in the Hamiltonian framework by the Dirac-Bergmann algorithm applied to $\widetilde{L}$ \cite{Dirac,Sund}. The Hamiltonian and the two first class constraints are:
\besub
\begin{align}
\label{e10}
\mH&  =  \frac{N}{2} g^{\alpha\beta} (p_\alpha - e A_\alpha) (p_\beta - e A_\beta) + \frac{m^2}{2}N  + p_N \dot{N}\,,
\\[6pt]
\label{e11}
\phi_1&=p_N\approx 0,\qquad \phi_2= g^{\alpha\beta}(p_\alpha-eA_\alpha)(p_\beta-eA_\beta)+m^2{ \approx 0} \,.
\end{align}
\label{e11-12}
\esub
Using the canonical equations  of the motion for  $x^\alpha$ and $p_\alpha$  and the condition \eqref{e3}, one obtains  that
\beq
\label{e13}
 \dot{I} = \frac{\partial I}{\partial \tau}+\{ I,\mH\}=N\omega \phi_2\approx 0\,.
\eeq
The total derivative of $I$ is weakly zero in the Dirac sense \cite{Dirac,Sund}, i.e., vanishes whenever $\phi_2$ does. Linear-in-the-momenta quantities with this property were called by  Kucha\u{r} \cite{Kuchar} conditional symmetries. Our  (\ref{e4}) and (\ref{e9}), extend  this notion to explicit dependence on the parameter $\tau$.

\par
To sum up, to  any proper conformal vector  which preserves also the  electromagnetic background is associated a conserved charge.  However for general  parametrisation  (including the  affine one),   the  latter have a seemingly  non-local contribution, which becomes local  for  trajectories with vanishing mass, $m=0$.

\section{Conformal symmetries  of pp-waves}
\label{Confpp}

In this section we apply the  procedure outlined  above to pp-waves, and show that the charges related  to conformal fields become local. Adopting the  terminology of Ref. \cite{b6},  we present the pp-wave metric in Brinkmann coordinates,
\beq
\label{e17}
g=d\bx^2+2dudv+H(u,\bx) du^2\,,
\eeq
where $\bx=(x^1, x^2)$.
Switching off  the electromagnetic field, $A_\mu=0$, we consider the extended Lagrangian $\widetilde{L}$ \eqref{e8}. The equations of the motion  are
\begin{subequations}
\label{eulgen}
\begin{align} \label{xeul}
{E_i}
&\equiv\ddot{x}^i -\frac{1}{2}\partial_i H(u,x)\dot{u}^2 -\frac{\dot{N}}{N}\dot{x}^i
= 0,
\\
\label{ueul}
{E_u}
&\equiv\ddot{u} - \frac{\dot{N}}{N}\dot{u} =0,
\\
\label{ueul2}
{E_v}
&\equiv\ddot{v} +\partial_i H(u,x) \dot{x}^i \dot{u}+\frac{1}{2} \partial_u H(u,x)\dot{u}^2 -\frac{\dot{N}}{N}\dot{v}=0,
 \\ 
\label{Neul}
 {E_N}
&\equiv H(u,x) \dot{u}^2+ 2 \dot{u}\dot{v} + \delta_{ij}\dot{x}^i \dot{x}^j + N^2 m^2 =0
\end{align}
\label{expliciteqs}
\end{subequations}
The $u-$equation \eqref{ueul} is the simplest one as it does not contain the profile $H(u,x)$ explicitly. The canonical momenta read
\beq
\label{expmomexactGW}
p_i = \frac{1}{N} \delta_{ij}\dot{x}^j,
\qquad
p_u =  \frac{1}{N} (\dot{v}+ H(u,x)\dot{u}),
\qquad
p_v =  \frac{\dot{u}}{N}\,,
\eeq
while $p_N$ is weakly zero.
The profile $H(u, x)$ does not depend on $v$, therefore its conjugate momentum $p_v$ is a constant of the motion associated with the covariantly constant vector $\p_v$. Note that for  a massive particle  the relation $p^\mu p_\mu =-m^2$ implies $p_v\neq 0$ as well as that
\beq
\label{dotvar}
\dot{v} = -\frac{1}{2\dot{u}} \Big((\dot{x}^i)^2 + H(u,x) \dot{u}^2 + m^2 N^2 \Big)\,,
\eeq
which is the algebraic solution of \eqref{Neul} with respect to $\dot{v}$.  Using the last equation in (\ref{expmomexactGW}), the non-local part of the integral of the motion takes the form
\beq
\label{nonlocalint}
\int\!\!\omega(x(\tau))N(\tau) d\tau=
\frac{1}{p_v}\int\!\!\omega(x(\tau))\dot{u} d\tau\,.
\eeq
\par
For pp-wave spacetimes this integral  can be calculated explicitly, both for the type N or  type O (conformally flat) cases. Let us recall some facts about conformal symmetries of a pp-wave space-time  \cite{b8}.  The components of a general conformal Killing vector are:
\begin{subequations}
\label{zetackv}
\begin{align}
  & Y^u  = \frac{\mu}{2}  \delta_{ij} x^i x^j + a_i(u) x^i + a(u) \\ \label{zeta2}
  & Y^v  = - \mu v^2 + \left( x^i a_i'(u)+ 2 b(u) - a'(u)\right) v + M(u,x,y) \\
  & Y^{i} = - \left( \mu x^i + a_i\right) v + \gamma_{ijkl} a_j'(u)x^k x^l + b(u) x^i - \epsilon_{ij}c(u)x^j  + c_i(u)
\end{align}
\end{subequations}
where $\mu$ is a non-zero constant and the ``prime'' $(\,\cdot\,)^{\prime}$ refers to $d/du$.
The corresponding conformal factor is
\begin{equation} \label{omega}
  \omega = \omega(u,x^i,v)= b(u) + x^i a_i'(u) - \mu v\,.
\end{equation}
The function $M(u,x,y)$ in \eqref{zeta2} satisfies suitable consistency conditions \cite{b8} while for $H(u,x,y)$ the relation
\begin{equation}
\label{rulH}
 \left[ \mu x^i +a_i(u)\right] \partial_i H = 2 \mu H + 2 a_i''(u) x^i -2 a''(u)+4 b'(u)
\end{equation}
must hold with $\mu=\const$
\footnote{ $\delta_{ij}$ is the Kronecker delta,
$\epsilon_{ij}$ the Levi-Civita symbol ($\epsilon_{12}=+1$) and $\gamma_{ijkl} = \frac{1}{2}\delta_{ij} \delta_{kl} - \epsilon_{ik} \epsilon_{jl}$.}.
  
Firstly, we observe that the integral \eqref{nonlocalint} can be computed explicitly for conformal vector fields which are ``chrono-projective'' \cite{Conf4GW,5Chrono},  defined by the property
\beq
\mL_Y \partial_v = \psi\partial_v
\where \psi =\const
\label{chronocond}
\eeq
$\psi$ here is the so-called chrono-projective constant. This condition is satisfied, e.g., by
 N-type null fluids and by exact plane gravitational waves \cite{Conf4GW}.

Chrono-projectivity  yields a relation between the conformal factor and the $u$-component of the  vector field,  $\omega \equiv \omega (u) = \frac{(Y^u)' - \psi}{2}$ ; see \cite{Conf4GW} for an explicit calculation.

Leaving exact gravitational waves to Section (\ref{ConfNflatEx})
we consider N-type null fluid pp-wave space-times \cite{b6}. The general form of the conformal Killing vector is 
\vspace{-1mm}
\begin{subequations}
\label{keatunfluidc}
\begin{align} \label{keatunfluidc1}
\omega(u)& = \frac{1}{2}(a'(u)-\psi),
\\
Y^u &= a(u),
\\
Y^i &= \omega(u) x^i + c_i(u) + \gamma \epsilon_{ij}x^j,
\\
Y^v &= -\psi v + \frac{a''(u)}{4}\bx^2 + c'_i x^i +E(u),
\end{align}
\end{subequations}
Since the conformal factor $\omega$ is a function of $u$ only, the integrand  in eq.  \eqref{nonlocalint} is a total derivative, allowing us to calculate the integral. We end up with a \emph{local} conserved charge
\beq
I = Y^u \left(p_u + \frac{m^2}{2p_v} \right) + Y^v p_v + Y^i p_i - \frac{m^2}{2p_v} \psi u\,.
\label{intchargenfa}
\eeq
$\psi=\const$ implies, in particular, that the last term is linear in $u$.
Note also the shift  $p_u \to p_u + \frac{m^2}{2p_v}$ in the momentum.  

Using the definition of the canonical momenta (\ref{expmomexactGW}), the quantity (\ref{intchargenfa}) can be expressed  in terms of the velocities. Interestingly, \emph{substitution of  (\ref{dotvar}) cancels the mass terms} in (\ref{intchargenfa}). Thus the conserved charge for the massive geodesic (\ref{intchargenfa}) \emph{coincides} with the one in the massless case. This statement holds for chrono-projective conformal fields.  

The case of type O pp-waves is slightly more involved. However, when the conformal factor is a function of $u$ only, the same arguments as for the type N apply, since \eqref{keatunfluidc1} still holds~: the conformal factor is a total derivative with respect to $u$ and the integral \eqref{nonlocalint} can be carried out. 

We are thus left to study fields as in \eqref{zetackv} which lead  to  conformal factor  $\omega= x^i a_i'(u) - \mu v$.
For simplicity, we consider the affine parametrisation with $N=1$ (\ref{affinecond}).  The last eqn. of \eqref{expmomexactGW} implies that  $u$  is proportional to the affine parameter (seen directly from \eqref{ueul}), and (\ref{nonlocalint}) becomes $\frac{1}{p_v}\int\!\!\omega(x(u)) du$ in this case. 

Let us stress that $p_v$ is a constant  of the  motion and  it can, if necessary,   be rewritten explicitly  in terms of the velocity $p_v=\frac{ d u}{d \sigma}$ where the parameter  $\sigma$ is defined by \eqref{affinecond}.

For any conformally flat pp-wave the profile  $H$ is (up to a coordinate transformation) a homogenous function of degree two in the transverse directions \cite{b8} and therefore the homothetic vector field
\beq
Y_{h} = x^i\partial_i + 2 v \partial_v
\label{homvfmg}
\eeq
is conformal with $\omega=1$.  The  corresponding  integral of the motion is
\beq
I_{h} = p_vx^i{x^i}'+2 v p_v +\frac{m^2}{p_v} u\,.
\label{mchargehom}
\eeq
Since all  pp-waves of the type O   are homogenous  of degree two,  the following  integral, along the trajectory,  can be computed  explicitly
\beq
\int v du= \int (\frac 12 \bx \cdot \bx'-\frac{m^2u}{2p_v^2}+\frac{I_h}{2p_v})du = \frac 1 4 \bx ^2-\frac{m^2}{4p_v^2} u^2+ \frac{I_h}{2p_v} u\,.
\eeq
Now, by virtue of \eqref{omega} the $v$  variable enters the  conformal  factor  at most linearly.  Therefore,  the integral in \eqref{nonlocalint} can be computed, as can also the conserved charge.

There remains the case when the conformal factor for  O type pp-waves  is of the form $\omega= {a_i}'x_i$ (cf. \eqref{omega}).  Let us note that \eqref{rulH}  implies that
$\ba= (a_1, a_2)$ obeys  the same equation as $\bx$ does (see \eqref{xeul} for  $N=1$). Along the trajectory we have therefore
\beq
{\bf a}'\cdot \bx=\frac 12 ({\bf a} \cdot \bx+u\bx\cdot  {\bf a}'-u{\bf a}\cdot  \bx')'
\eeq
and  the integral of $\omega$ in \eqref{nonlocalint} can again be explicitly computed.

In summary, for N-type pp-waves, the integrals of the motion can be computed in a straightforward manner ; for O-type pp-waves, the associated integrals of the motion can be computed explicitly  in the affine parametrisation. This result  can be extended to an arbitrary parametrisation, considered in sec. \ref{ModConfpp}.

\par
Several questions can now be asked: do  these conserved  charges  provide  new information and  can they be identified with Noether symmetries (as for  Killing fields),  or  are they more  general  non-point   symmetries~?

Next, can we find  electromagnetic backgrounds preserved by the conformal fields and  derive  corresponding integrals of the  motion?  What is the meaning, in this context, of the  Eisenhart-Duval lift ?  We  turn to these questions in the following  sections.

\section{Further aspects of integrals of the motion for pp-waves}\label{furtherStud}

\subsection{Eisenhart-Duval lift and massive geodesics}\label{EDlift}

According to the  Eisenhart-Duval (E-D) framework   \cite{Eisenhart,Bargmann,DGH91,BekaertMorand,dissip}, the  non-relativistic dynamics in two space dimensions  emerges  from the light-like reduction of null geodesics in a $3+1$ dimensional relativistic space-time  (\ref{e17}). Referring to the literature  for details, we just sketch the main idea. In the $u$-parametrisation  the geodesic equations  \eqref{eulgen} can be rewritten  as,
\vspace{-3mm}
\besub
\begin{align}
\label{e18a}
{\bx}''&=\frac {{\vnabla} H}{2},
\\
\label{e18b}
 v'&=-\frac{1}{2}\left({\bx'}^2+H\right)-\frac{m^2}{2 p_v^2}\,.
\end{align}
\esub
The transversal part of the geodesic equations decouple and can be considered  as the Newton equations with the   $u$-coordinate viewed as non-relativistic time. The non-relativistic motions lifted to motion along geodesics with  fixed $m$ and $p_v$ . (In the original approach, $m=0$ and $p_v$ is identified with the non-relativistic mass \cite{Eisenhart,Bargmann,DGH91}.)

For massless geodesics the integrals  of the motion associated with conformal fields are local cf. \eqref{e4aff} and \cite{CGHHZ,Conf4GW}. Moreover, under some assumptions, these integrals project  onto  integrals of the motion for the underlying non-relativistic system.

In the previous section, we have noted that the conformal vector fields of N-type pp wave space-times  satisfy the chrono-projectivity condition \eqref{chronocond}.
  Therefore, only their $v-$component  depends on $v$ according $Y^v = Y^v_1(u, \bm{x}) - \psi v$, see \cite{Conf4GW} and references therein.

The $v$-coordinate can be expressed
 in a non-local form in terms of the  \emph{action integral} of the projected non-relativistic dynamics,
\begin{eqnarray}
\label{e19}
v=-\int \frac{1}{2}\Big({\bx^{\prime}}^2+H(u, x)+\frac{m^2}{p_v^2}\Big)du
= -{}{{}}\int{L_{NR}}\ du - \frac{m^2u}{2p_v^2}+\const\, ,
\end{eqnarray}
where $L_{NR}$ is the non-relativistic Lagrangian. (For pp-waves of type O $v$ can be  obtained directly and consequently be eliminated).

The projected integral of the motion behaves similarly, as  illustrated by  the Kepler problem \cite{KHarmonies, Zhang:2019koe}.
It is  worth to notice that  the constraint $p_\mu p^\mu = -m^2$ in \eqref{e11} allows us to express $p_u$ as
\beq
p_u = -\frac{p_i^2}{2 p_v} + \half H(u,x^i) p_v - \frac{m^2}{2p_v}\,,
\label{NRHwb}
\eeq
which is (up to a constant)  minus the non-relativistic Hamiltonian $H_{NR}=-p_u$ obtained by projection to transverse space.

The transverse eqs.  \eqref{e18a}  are identical for both the massive and massless geodesics; the only difference is that  in the massive case the  $v$ coordinate is shifted by the linear-in-$u$ term as in \eqref{e19}. In the massless case $m=0$ \eqref{e19}  reduces to the horizontal lift \cite{Eisenhart,Conf4GW}.  On the other hand we know how the $v$ variable enters  the  conformal fields \eqref{zetackv} and consequently  the integrals of the motion associated with them (see the previous section).  Thus we can  find the difference  between the massive and massless charges  expressed in terms of  the variables $u,\bx $ only.
\par
Let us start with  the conserved  charges for a massive geodesic of a type N  pp-wave. Using eq. \eqref{e19}  $v$ can be eliminated ; then the 4-dimensional integral of the motion \eqref{intchargenfa} projects downstairs as,
\beq
I=p_v\!\left(-\frac{1}{2}({\bx^{\prime}}^2-H(u,\bx))Y^u(u)+{\bf Y}(u,\bx)\cdot \bx^{\prime}+Y^v_1 (u,\bx)+{\psi}\int \!\!L_{NR}\  du \right).
\label{redI}
\eeq

The terms  which contain  $m$  reduce to a constant and we recover the  charge for massless geodesics; expressed in terms of  $u,\bx$. However, for a proper chrono-projective transformation ($\psi \neq 0$) the integral term  is non-local, unlike the expression  \eqref{intchargenfa} in full four-dimensional  space-time.

The same situation holds for  any   conformal vector  field of the  pp-wave.  Using the form of $Y$  given in \cite{b6}, (especially eqn. \#  (12))  one finds  that after eliminating the $v$ coordinate, the terms with mass reduce  to  a constant.  Thus the  conserved charge  coincides with the massless one, expressed  in terms of $\bx$ and $u$ -- just like the transverse parts of  the  geodesic equations for the pp-wave are  identical for both massive and massless  geodesics.

\goodbreak
\subsection{Conformally related metrics}

Let us consider a pp-wave metric $g$  and a new one $\hat g$ conformally related to $g$,
\beq
\hat  g= \Omega^2(u)\,g\,,
\eeq
where $\Omega(u)$  is an arbitrary function. Then $\hat  g$ has  the same conformal algebra as $g$; however, the Killing, homothetic and proper conformal transformations can be different \footnote{These  metrics  are physically  inequivalent. E.g., one  can be a vacuum solution, the other  not.}.  On the other hand, after a suitable transformation to new coordinates $\tilde u,\tilde{\bx} ,\tilde v$, the metric  $\hat g$ takes the pp-wave form \cite{b6}.   As  we have seen above,  the integrals of the motion in a pp-wave for  massive  geodesics coincide with  those of the massless ones when  expressed in terms  $u,\bx$. 
However thosee integrals for $g$ and $\hat g$, expressed in terms of $u$ and $ \bx$,  coincide up to a constant (this  can be  checked  also directly using eqn. \# (12) of \cite{b6}).
Thus for both, physically inequivalent, pp-waves  the integrals of the motion  associated with the conformal generators  can be directly related.
\par
Let us  illustrate  this observation  with the  proper conformal transformations of a particular type of space-time which preserve, in addition,  a given electromagnetic background.

In the Minkowski space-time there are four proper conformal transformations, see section \eqref{MinkEx} .
For example, we have  the so-called standard special conformal vector (In the non-relativistic context, $\p_u$ is time translation and $Y_K$ is an expansion \cite{Conf4GW}), see eqn.  \eqref{expansionM}  below.
 In order to obtain geodesically complete metrics, we combine  $Y_K$ with the Killing vector $\partial_u$,
\beq
\label{e16}
Y^{(1)}=Y_K+\epsilon^2\partial_u=  (u^2+\epsilon^2)\partial_u-\frac1 2 \bx^2\partial _v+u{\bf x }\cdot \vnabla, \qquad \omega(u) =u\,.
\eeq

We  focus our attention on gravitational or/and electromagnetic backgrounds. For example,  in refs.   \cite{b7,b3,b4,b5}  some  electromagnetic as well as gravitational fields which satisfy \eqref{e3} were  studied. Due to the modification \eqref{e16}   these backgrounds  form  non-singular  pulses.
Below we give further examples of gravitational fields for which  $Y^{(1)}$ is  a conformal vector, and more electromagnetic backgrounds which are preserved by this field.

Turning around the question, we ask which pp-wave space-times do admit $Y^{(1)}$ as conformal vector with identical conformal factor $\omega=u$. One finds that this happens  iff  its profile $H$ satisfies
\beq
\label{e23}
(u^2+\epsilon^2)\partial_ uH+u\,\bx\cdot \vnabla H + 2uH=0\,.
\eeq
The solution of this equation  is of the form
\beq
\label{e24}
H=\frac{2\epsilon^2}{u^2+\epsilon^2}K\big(\frac\bx{\sqrt{u^2+\epsilon^2}}\big)\,,
\eeq
where $K$  is an arbitrary function of two variables. Thus for this family  of pp-waves   the  corresponding integral  of the motion are \eqref{e4aff} (see also \eqref{redI}),
\beq
\label{e25}
I^{(1)}= p_v\left( -\frac{1}{2}\bx^2+u\bx\cdot{\bx}^{\prime}+\frac{1}{2}(-{\bx^{\prime}}^2+H)(u^2+\epsilon^2)-\frac{m^2}{2p_v^2}\epsilon^2\right),
\eeq
where the last term is just a mass-dependent constant.

To illustrate our observation  let us now rewrite \eqref{e24} as
\beq
\label{e26}
H=\frac{2\epsilon^2}{(u^2+\epsilon^2)}\left (\widetilde{K}(\frac\bx{\sqrt{u^2+\epsilon^2}})+\frac{\bx^2}{2(u^2+\epsilon^2)}\right)\,,
\eeq
where $\widetilde{K}$ is arbitrary function. Then \eqref{e25} becomes
\beq
\label{e27}
I^{(1)}=-\frac{\epsilon}{2}(\rho{\bx}^{\prime}-{\rho}^{\prime}\bx)^2+\epsilon^2\widetilde{K}(\frac{\bx}{\sqrt{\epsilon}\rho})-\frac{\epsilon^2m^2}{2p_v^2}
\quad\text{where}\quad
\rho=\sqrt{u^2+\epsilon^2}/\sqrt{\epsilon}\,.
\eeq
On the other hand, consistently with the general theory \cite{b14,b15} (and verified also by direct calculation), the Niederer transformation \cite{b13}
\beq
\label{e28}
u=\epsilon\tan(\tilde u), \quad
\bx =\frac{\epsilon \tilde \bx}{\cos(\tilde u)}\,,
\\
\eeq
 relates the transverse part of  the geodesic equation to  a ``time"-independent set  of equations,
\beq
 \label{e29}
 \bx''=\widetilde \vnabla\widetilde K(\tilde \bx)\,.
\eeq
 Next,	 let us note that  the metric $\tilde{g}$  defined by the profile $2\widetilde{K}$ is conformally related  to the metric $g$, defined by \eqref{e26}. Indeed,  supplying  the  transformation \eqref{e28} by
$v=\epsilon\tilde v-\epsilon\tan(\tilde u){{\bf\tilde  x}^2}/{2}$
one has
\beq
 g = \frac{\epsilon^2}{\cos^2(\tilde u)}\left(2\widetilde{K}(\tilde \bx) d\tilde u^2+2d\tilde u{}d \tilde v+d\tilde \bx^2\right)= \frac{\epsilon^2}{\cos^2(\tilde u)}\,\widetilde g\,.
\eeq
In the  new variables the vector  \eqref{e16}   takes the form  $\epsilon\partial_{ \tilde u}$  and its conformal  factor is  $\epsilon \tan(\tilde u)$.  On the other hand,   $\epsilon\partial_{ \tilde u}$   is a Killing vector  of $\widetilde g$ for which the suitable integral of the motion  \eqref{e4aff}  can  easily be obtained; it turns out to be  the energy for the projected dynamics,
\beq
\widetilde I =\frac{1}{2}\tilde \bx'^2-\widetilde{K}(\tilde \bx) =E\,,
\eeq
corresponding to  eqs. \eqref{e29}.
Finally, the integral $I^{(1)}$  in the new variables takes the form
\beq
\label{e31}
I^{(1)} =-\epsilon^2\left(\widetilde I+\frac{m^2}{2p_v^2}
\right)\,.
\eeq
Thus, in full agreement  with  our  general observations, in the massive  case  there is only a constant  between projected and lifted integrals of the motion. The   charges for  both conformally related metrics $g$ and $\tilde{g}$ coincide.

\subsection{Conformally invariant electromagnetic backgrounds}\label{cinvem}

As  mentioned already, if a Killing vector preserves the  electromagnetic background, see  \eqref{e3}, then one can  construct a suitable integral of the motion (see e.g. \cite{Ilderton} for a detailed  discussions).
The question is whether one  can find a  pp-wave  admitting a conformal field which preserves some  electromagnetic backgrounds and the corresponding charge localizes in affine parametrisation. We give here some examples of such a situation (extending some electromagnetic vortices, see \cite{b5}  for further discussion).
\par
Let us consider the pp-wave space-time defined by  the profile \eqref{e24} (in  particular, the Minkowski space-time). Then the field $Y^{(1)}$, eqn. \eqref{e16},    is a conformal one. Now we take  an electromagnetic field,
$ 
\mA=A_udu= A(u,\bx)du\,,
$ 
 where $A$ is an arbitrary function. Then  $p_v$ is again   a constant of the  motion and $u$  is proportional to the affine  parameter.  Let us  impose the condition \eqref{e3}, i.e.   we assume that  the  potential is preserved by  $Y^{(1)}$ up to a gauge transformation $\phi$.
Straightforward computations  imply that $\phi$  is a function of $u$ only; however, then   one can find   a suitable gauge transformation  of the electromagnetic potential such that $\phi=0$; thus,  without loss of generality,  we can  assume this condition  (for a fixed field $Y$ such a choice of $\phi$ is always possible; however, not necessary explicitly given).  Then  eqn.  \eqref{e3}  imposes only  one condition   on  the profile $A$, which is, remarkably,  of the  same form as \eqref{e23} (after the substitution $H\rightarrow A$). Thus   we obtain a whole family (cf. eqn.   \eqref{e24})  of electromagnetic profiles  which are preserved by $Y^{(1)}$.
Of course,  one  can put  $A  =H$ (cf.  the double copy conjecture \cite{b17}); however,  $A$ and $H$  can be chosen independently (for example, we can take  $H=0$  i.e.  Minkowski space-time).
 For  such  pp-waves and electromagnetic fields the  integral of the motion associated with  $Y^{(1)}$  can be written down explicitly, see  eqn. \eqref{e4aff},
\beq
\label{e35}
I^{(1)}_A=I^{(1)} +\frac{e}{p_v}(u^2+\epsilon^2)A\,,
\eeq
where $I^{(1)}$ is given by (\ref{e25}).
\par  Let us note finally that such an integral of the motion can bring some new information even for the Minkowski space-time. Indeed, the  electromagnetic fields constructed are  preserved by $\partial_v$ which leads to charge  $p_v$; however,  symmetries related to  other Killing fields (Poincar\'e generators) are in general broken and do not provide integrals of the motion (see also the discussion in sec. \ref{lcgs} as well as in \cite{b5}).

\section{Distorted conformal symmetries}
\label{ModConfpp}

In this section we first review some general aspects of the Noether symmetry approach \cite{Olver,Bluman,Ibra} and explain how the previously found integrals of the motion are related to symmetry transformations of the action. Our main statement is that modifying the standard procedure  allows us to derive the generators of the conserved charges by a mass -dependent ``distortion'' of the conformal Killing vectors. The connection of these charges to a more general symmetry is also established and their general properties in phase space are investigated.

\subsection{The Noether symmetry approach to  geodesic systems}\label{NoetSymmSec}

Let us first recall some basic facts   \cite{Olver,Bluman,Ibra}. A transformation is a Noether symmetry if it leaves the action integral form-invariant up to a surface term. We use the Lagrangian \eqref{e8} as a model to illustrate the basic properties of the general theory. In infinitesimal form  the symmetry generator is
 \begin{equation} \label{Xgen}
  X= \chi \frac{\partial}{\partial \tau} + \Upsilon^N \frac{\partial}{\partial N} + \Upsilon^\alpha \frac{\partial}{\partial x^\alpha}\,.
\end{equation}
Its extension to the space of the first derivatives ,
\begin{equation}
\label{prolongation}
  \mathrm{pr}^{(1)}X = X + \big(\frac{d \Upsilon^{\alpha}}{d\tau} - \dot{x}^\alpha \frac{d \chi}{d \tau}\big) \frac{\partial}{\partial \dot{x}^\alpha}\,
\end{equation}
(called the first prolongation of $X$)
is required to satisfy the infinitesimal invariance criterion,
\begin{equation}
\label{critinv}
\medbox{
  \mathrm{pr}^{(1)}X (L) +L \,\frac{d \chi}{d \tau} =
 \frac{d F}{d\tau}\,}
\end{equation}
for some function $F$.

 Equation \eqref{critinv} is our starting point for  searching for symmetries. For a given Lagrangian $L$ one tries to find appropriate vectors $X$ and corresponding functions $F$ which satisfy \eqref{critinv}. The  conserved quantity associated with a symmetry generator $X$ is,
\begin{equation}
\label{genQ}
\medbox{
  Q = \Upsilon^\alpha \frac{\partial L}{\partial \dot{x}^\alpha} + \chi \left(L- \dot{x}^\alpha \frac{\partial L}{\partial \dot{x}^\alpha} \right) - F \,. }
\end{equation}
(A  term $\Upsilon^N \frac{\partial L}{\partial \dot{N}}$ could also be included in  \eqref{prolongation}, but for the Lagrangian $\widetilde{L}$ of \eqref{e8} such additions are trivial and are therefore omitted.)

The simplest case is to consider \emph{Noether  point symmetries}, where the generators do not dependend on the derivatives. That is, we can have at most~:
\begin{equation}
\label{gencoeff}
  \chi = \chi (\tau, N, x), \quad \Upsilon^N = \Upsilon^N  (\tau, N, x), \quad \Upsilon = \Upsilon^\alpha  (\tau, N, x)\,.
\end{equation}
Such a case can be treated algorithmically since the coefficients of the derivative terms ($\dot{N}$, $\dot{x}$ and their powers) appearing in \eqref{critinv} must be zero.
 In this manner a set of over-determined linear, partial differential equations is obtained. Their solution (if it is not trivial, i.e. $X=0$, $F=\const$) yields a {Noether point symmetry}.

 On the other hand, if the coefficients in \eqref{Xgen} are allowed to depend also on derivatives of the configuration space variables, e.g. $\chi=\chi(\tau,N,x,\dot{N},\dot{x},...)$ etc., then the previous procedure cannot be followed and \eqref{critinv} has to be treated in its totality as a single master equation. For this reason, non-point (or generalized) symmetries are much more difficult to be extracted.

For the geodesic problem, the list of Noether point symmetries is well known \cite{Prince,Hojman1,Andr1,Andr2,Hussain2}. In the massive case the basic Noether symmetry generators are related to the homothetic algebra of the metric \cite{Prince,Andr1,Andr2}. For null geodesics however, the results can be extended to include all conformal Killing fields, $Y$, \cite{Prince,Katzin}. (See also \cite{Conf4GW} for a different approach). In Table \ref{tab1} we collect the known results on point symmetries  for the geodesic action both for the affinely parametrised case $L_0= \frac{1}{2}g_{\alpha\beta}\frac{d x^\alpha}{d\sigma}\frac{d x^\beta}{d\sigma}$ and for the parametrisation-invariant Lagrangian $\widetilde{L}$ \cite{tchris}. In the former case there is no $N$ field, while in the latter the Einbein field is considered as a degree of freedom on equal footing with $x^\alpha$.

\begin{table}
\centering
\begin{tabular}{ |c|c|c||c|c|c|c||c|}
 \hline
 \multicolumn{3}{|c||}{$L=L_0:=  \frac{1}{2}g_{\alpha\beta}\frac{d x^\alpha}{d\sigma}\frac{d x^\beta}{d\sigma} $}  & \multicolumn{4}{c||}{$L=\widetilde{L}:= \frac{1}{2N}g_{\alpha\beta}\dot{x}^\alpha \dot{x}^\beta- N\frac{m^2}{2}$} & \multicolumn{1}{c|}{Geometric conditions} \\
 \cline{1-7}
 $\chi(\sigma,x)$ & $\Upsilon(\sigma,x)$ & $F(\sigma,x)$ & $\chi(\tau,N,x)$ & $\Upsilon^N(\tau,N,x)$ & $\Upsilon(\tau,N,x)$ & $F(\tau,N,x)$ & {on space-time vectors  $Y_0$, $Y_h$}  \\
 \hline
 const.  &  -    & const. & $\chi(\tau)$  & $\dot{\chi}(\tau) N$ & -    & const. & - \\
 \hline
 -  &  $Y_0(x)$  & const.  & -  & - & $Y_0(x)$  & const. & $\mathcal{L}_{Y_0}g_{\alpha\beta} =0$\\
 -  &  $\sigma Y_0(x)$  & $\Phi(x)$  & -  & - & -  & - & if also $Y_0^\alpha = \nabla^\alpha \Phi$ \\
 \hline
 $2 h \sigma$  &  $Y_h(x)$  & const. & -  & -&  -  & -   & $\mathcal{L}_{Y_h}g_{\alpha\beta} =2 h g_{\alpha\beta}$, $h=$const. \\
 $h \sigma^2$  &  $2 \sigma Y_h(x)$  & $\Sigma(x)$  & -  & -&  -  & -   & if also $Y_h^\alpha = \nabla^\alpha \Sigma$ \\
 \hline
\end{tabular}
\caption{All point symmetry solutions of \eqref{critinv} for  affinely parametrised and resp.  parametrisation-invariant Lagrangians.} \label{tab1}
\end{table}

In the affine case -- apart from the trivial time translation which implies that the Hamiltonian is a constant -- the Killing and homothetic fields, $Y_0$ and $Y_h$ respectively, contribute Noether point symmetries. What is more, both can be used to  provide an additional conserved charge when they happen to be gradient vectors \cite{Andr1}. On the contrary, for the parametrisation invariant Lagrangian the generator for arbitrary transformations in time ($\chi(\tau)$ remains an arbitrary function) and the integrals of the motion generated by Killing vectors $Y_0$ are obtained \cite{tchris}.

Let us turn to the pp-wave space-time \eqref{e17} and consider a free particle of mass $m$ described by the Euler-Lagrange equations \eqref{expliciteqs} (with $A_\alpha =0$).  As we noted in sec. \ref{Confpp} the localization of charges is  related to the fact that  on shell the conformal factor $\omega$ is the \emph{total time derivative of a suitable function},
\beq
\label{derf}
\omega(x^\alpha(u))={( f^\alpha {x_\alpha'})}'\,
\eeq\vskip-7mm
\noindent{where}
\begin{subequations}\vspace{-2mm}
\label{coefef}
\begin{align}
  f^u  = & 0,
 \\
  f^v  = & \frac{1}{2} u  \left(x^i a_i'(u) - a' (u)+2 b(u)-2 \mu v \right) + \frac{1}{2} x^i a_i (u) + \frac{\mu}{4} \delta_{ij} x^i x^j
 + \frac{1}{2} a(u) - \frac{m^2}{p_v^2} \frac{\mu}{4} u^2\,,
  \\
f^{i} = & -\frac{1}{2}u \left(\mu\, x^i + a_i(u) \right)\,.
\end{align}
\end{subequations}
Then along trajectories we have see \eqref{eulgen} and \eqref{expmomexactGW}
$$
f^\alpha p_\alpha=\frac{f^\alpha\dot x_\alpha}{N}=f^\alpha x'_\alpha\frac{\dot u}{N}
\;\Rarrow\;
\frac{d}{d\tau} { (f^\alpha p_\alpha)}=
\frac{\dot u }{N} (f^\alpha x_\alpha')'\dot u =(f^\alpha x_\alpha')'p_v^2 N=\omega p_v^2 N.
$$ \vspace{-3mm}
Thus we obtain
 \beq
\label{fe}
 m^2\!\int\!\!\omega(x(\tau))N(\tau) d\tau=\frac{m^2}{p_v^2}f^\alpha p_\alpha  + \const = \frac{m^2}{N p_v^2}f^\alpha    \dot x_\alpha  + \const
 \eeq
The \emph{charge is local}.

Note that \eqref{fe}  is valid in any arbitrary parametrisation (because the momenta are parametrisation invariant).
Therefore the integral of the motion  $I$   given by \eqref{e9} is equivalent to \vspace{-4mm}
\begin{equation}\label{netint}
  Q = \Upsilon^\alpha \frac{\partial \widetilde{L}}{\partial \dot{x}^\alpha}\,,
\end{equation}
associated with the modified (``distorted'') vector field
\beq
\label{upsdef}
\medbox{
\Upsilon^\alpha = Y^\alpha + \frac{m^2}{p_v^2} f^\alpha\,.}
\eeq
We now illustrate the new formulation by examples:
\begin{itemize}
\item Homothety~: the comparison of  \eqref{homvfmg}  with  eqn.  \eqref{zetackv} yields
\begin{subequations} 
\begin{align}
Y^u = 0, \quad &\implies \quad \mu = a_i = a(u) = 0 \\
 Y^v=2v \quad &\implies \quad b=1, \ M = 0 \\
 Y^i = x^i \quad &\implies \quad \gamma_{ijkl} = \gamma(u) = c_i(u) =0,
\end{align}
\end{subequations}
with $\omega = b =1$. Eqns. \eqref{coefef} give
$
f^u = f^i = 0, \, f^v =  u.
$
Then  \eqref{upsdef} used in \eqref{netint} implies the integral of motion \eqref{mchargehom}, as expected.
\item pp-waves of the type N: from \eqref{keatunfluidc}  we get
\begin{subequations}
\begin{align}
Y^u& =  a(u)  \quad \implies \mu = a_i =0, \\
\omega(u)&=  \frac{a'(u)-\psi}{2} \quad \implies b(u) = \frac{a'(u)-\psi}{2},\\
 Y^v&= -\psi v + \frac{a''(u)}{4}\bx^2 + c'_i x^i +E(u)  \implies   M = \frac{a''(u)}{4}\bx^2 + c'_i x^i +E(u) \\
 Y^i &= \omega(u) x^i + c_i(u) + \gamma \epsilon_{ij}x^j \implies \text{automatically satisfied}.
\end{align}
\end{subequations}
Thus \eqref{coefef} is solved by
$
f^u = f^i = 0, \, f^v = \frac{1}{2}\left(Y^u - u\psi\right) ,
$
which results through \eqref{netint} to a conserved charge consistent with \eqref{intchargenfa}.
\end{itemize}
By comparing \eqref{genQ} with $\chi=F=0$ and $L=\widetilde{L}$ derived from the Noether symmetry approach with the conserved quantity in \eqref{netint} obtained with no reference to Noether symmetry suggests that the latter is generated by the modified vectorfield $\Upsilon$ in \eqref{upsdef}, which plays a role similar to a Noether point-symmetry. We will return to this point later.
However it can be easily checked that \eqref{upsdef} fails to satisfy the symmetry criterion \eqref{critinv}. It is only in the $m=0$ case that  we recover what we know from the general theory, namely that for null geodesics the conformal Killing vectors $Y$ yield linear-in-the momenta conserved quantities \cite{Prince,Katzin}.

An intriguing observation is that the vectors $\Upsilon$ of \eqref{upsdef} do not necessarily close to an algebra. They trivially do so when $m=0$, where they reduce to the conformal Killing vectors of the metric.
This leads us to inquiring whether the Noether symmetry approach can be modified so that it   explains the form of the ``distorted'' vector \eqref{upsdef} when $m\neq 0$ and the emergence of the conserved charge \eqref{netint}. 

\subsection{Modification of the Noether approach and the role of the constraints}\label{Geosec}

In the previous subsection we outlined the procedure of deriving a point symmetry generator satisfying \eqref{critinv}. Take for example the quadratic parametrisation-invariant Lagrangian $\widetilde{L}$ of \eqref{e8} for space-time \eqref{e17}.
The  invariance  criterion  \eqref{critinv}
requires to solve the system of partial differential equations for $\chi$, $\Upsilon^N$ and $\Upsilon$ which demands the coefficients of $\dot{u}$, $\dot{v}$, $\dot{x}$, $\dot{y}$  to vanish. This  scenario  leads directly to the right part of Table \ref{tab1}, which, -- leaving out the parametrisation invariance -- is  equivalent to the Killing equations. Instead of following strictly this procedure, we choose to modify it in a manner that is consistent with the equations of motion.

First we notice that 
 it is possible to express the velocity $\dot v$ in terms of the remaining variables, see eqn. \eqref{dotvar}. Then we eliminate  $N$  by using  $N=\dot u/p_v$ which is the first integral of \eqref{ueul}, see the last of eq. \eqref{expmomexactGW}. Substituting into \eqref{critinv} and  collecting the coefficients of the remaining velocities $\dot \bx$ and $\dot u$  leads us to the following weaker conditions on $\Upsilon$. (For the rest we consider $\chi=\Upsilon^N=0$, $F=\const$).
  \vspace{-5mm}
\goodbreak
\begin{subequations}
\label{symtotal}
\begin{align}\label{sym1-4}
  & \frac{\partial \Upsilon^u}{\partial v} = 0, \quad \frac{\partial \Upsilon^1}{\partial y} + \frac{\partial \Upsilon^2}{\partial x} = 0, \quad \frac{\partial \Upsilon^u}{\partial x} + \frac{\partial \Upsilon^1}{\partial v} = 0, \quad \frac{\partial \Upsilon^u}{\partial y} + \frac{\partial \Upsilon^2}{\partial v} = 0, \\
   \label{sym5}
  & \frac{\partial \Upsilon^u}{\partial u} - \frac{m^2}{p_v^2} \frac{\partial \Upsilon^u}{\partial v} + \frac{\partial \Upsilon^v}{\partial v} - 2 \frac{\partial \Upsilon^2}{\partial y} = 0, \\ \label{sym6}
  & \frac{\partial \Upsilon^u}{\partial u} - \frac{m^2}{p_v^2} \frac{\partial \Upsilon^u}{\partial v} + \frac{\partial \Upsilon^v}{\partial v} - 2 \frac{\partial \Upsilon^1}{\partial x} = 0, \\
  \label{sym7}
  & (H-\frac{m^2}{p_v^2})\frac{\partial \Upsilon^u}{\partial y} + 2 \frac{\partial \Upsilon^v}{\partial y} + 2 \frac{\partial \Upsilon^2}{\partial u} - (H+\frac{m^2}{p_v^2}) \frac{\partial \Upsilon^2}{\partial v} =0, \\ \label{sym8}
  & (H-\frac{m^2}{p_v^2})\frac{\partial \Upsilon^u}{\partial x} + 2 \frac{\partial \Upsilon^v}{\partial x} + 2 \frac{\partial \Upsilon^1}{\partial u} - (H+\frac{m^2}{p_v^2}) \frac{\partial \Upsilon^1}{\partial v} =0,\\
  \nonumber
  & (H-\frac{m^2}{p_v^2}) \frac{\partial \Upsilon^u}{\partial u} + \frac{\frac{m^4}{p_v^4} - H^2}{2} \frac{\partial \Upsilon^u}{\partial v} + 2 \frac{\partial \Upsilon^v}{\partial u} - (H+\frac{m^2}{p_v^2})  \frac{\partial \Upsilon^v}{\partial v} + \Upsilon^u \frac{\partial H}{\partial u}  \\ \label{sym9}
  & + \Upsilon^1 \frac{\partial H}{\partial x} + \Upsilon^2 \frac{\partial H}{\partial y} = 0 ,
\end{align}
\end{subequations}
These equations differ from those satisfied by a conformal Killing vector only in terms which involve the mass, $m$. The ``distorted" $\Upsilon$ in \eqref{upsdef}  satisfies  the above set of equations; consequently, the modification of the Noether procedure by invoking  the known integrals of the motion \eqref{expmomexactGW}, \eqref{dotvar}
before collecting  coefficients, leads us to the desired fields.

As far as the conservation  of  $Q$ in \eqref{netint} is concerned, it is straightforward to show that
\begin{equation}
\label{nothidentity}
  \frac{d\cQ}{d\tau} = -2 N \omega_m E_N(\widetilde{L}) - \Upsilon^\alpha(u,v,x,y) E_\alpha(\widetilde{L}) + \frac{m^2}{N} \omega_m  \left(N^2-\frac{\dot{u}^2}{p_v^2}\right)\,,
\end{equation}
where $E_N(\widetilde{L})=0$, $E_\mu(\widetilde{L})=0$ are the Euler-Lagrange equations of \eqref{expliciteqs} and
\beq
\omega_m =  \omega - \frac{m^2}{2p_v^2} \mu\, u\,.
\eeq
From \eqref{nothidentity} we see that the right hand side provides us with  an additional condition which is satisfied on mass shell: the first integral of \eqref{ueul}, $\dot{u}= p_v N$, with $p_v$ a constant. The $\Upsilon$ defined by \eqref{upsdef} is a vector field on the  configuration space   only provided $p_v$ entering its right hand side  is viewed as a parameter. It does not  formally define a point symmetry; this is clearly seen from eqn.  \eqref{nothidentity} because the right hand side is not  just a combination of Euler-Lagrange equations. However, if we restrict ourselves to  trajectories  whose momentum conjugated to $v$ takes the value $p_v$  (entering \eqref{upsdef}), then the last term on the right hand side vanishes and  we obtain the desired conservation law.

Now an additional question arises: Can the $\Upsilon$  be related to some formal Noether symmetry that directly satisfies \eqref{critinv}~? The answer is affirmative~: we just need to eliminate the constants of the motion in  the vector \eqref{upsdef} by their velocity-equivalents on mass shell.
With this substitution in \eqref{upsdef} the new generalized vector $\Upsilon$ satisfies the symmetry criterion \eqref{critinv} on the mass shell ($\chi=\Upsilon^N=0$, $F=$const.) without the need to involve in addition the constraint equation or an integral of the motion. What is more, this  symmetry (which is now  a non-point Noether symmetry due to its dependence on velocities) \emph{can} satisfy \eqref{critinv}, not just for $L=\widetilde{L}$, but also for the affinely parametrised Lagrangian $L_0$.

The above procedure can be realized  also at the Hamiltonian level. Take the Killing tensor denoted by $K_{\mu\nu}=(\partial_v \otimes \partial_v)_{\mu\nu}=\delta_{\mu v}\delta_{\nu v}$. Then the system possesses the conserved charge $\mK=K^{\alpha\beta}p_\alpha p_\beta$. Let us promote the corresponding conservation law to a constraint,
\beq
\phi_3 \equiv  K^{\alpha\beta}p_\alpha p_\beta - \kappa   \approx 0 \,,
\label{phi3constr}
\eeq
where $\kappa$ is the constant value taken  by $\mK$ along a trajectory ($\kappa=p_v^2$ is used here so as to avoid confusion with seeing  $p_v$ as a variable on the phase space). Integrals of the motion viewed as first class constraints have  previously been studied in \cite{Pons} from the perspective of the gauge transformations they generate. 

The constraint \eqref{phi3constr} is consistent with the evolution of the system since it does not generate additional restrictions~: $\dot{\phi}_3 =\{\phi_3,\mathcal{H}\}=0$. It also commutes with both $\phi_1=p_N$ and $\phi_2=g^{\mu\nu}p_\mu p_\nu + m^2$ which makes it a first class constraint. Considering a quantity
which is linear in the momenta,
 $Q=\Upsilon^\alpha p_\alpha$, and imposing the condition $\dot{Q} \approx 0$; a conditional symmetry in  Kucha\u{r}'s sense emerges.
  Assuming  the additional  constraint $\phi_3$
 this condition  can be rewritten as,
\begin{equation}
\label{conscond}
  \begin{split}
    \dot{Q} =\{Q, \mathcal{H}\} \approx 0 \Rightarrow \dot{Q} & = \omega_m(x) N \phi_2 + \tilde{\omega} (x) N \phi_3 \\
    & =\omega_m N \left( g^{\mu\nu}p_\mu p_\nu + m^2\right) + \tilde{\omega} N K^{\mu\nu}p_\mu p_\nu - \omega_2 N \kappa\,,
  \end{split}
\end{equation}
where the multiplying factors on the right hand side are chosen to be consistent with what appears on the left. The left hand side is purely quadratic in the momenta, hence we demand $\tilde{\omega} = \frac{m^2}{\kappa}\omega_m $, which yields
\begin{equation}
\label{condsym2}
  \{Q,\mathcal{H}\}  = \omega_m N \left(g^{\mu\nu}+ \frac{m^2}{\kappa}K^{\mu\nu} \right)p_\mu p_\nu\,,
\end{equation}
cf. (\ref{e13}), leading subsequently to the geometric condition
\begin{equation}
\label{modconf}
  \mathcal{L}_\Upsilon g_{\mu\nu}  = {2}\omega_m  \left(g_{\mu\nu}+ \frac{m^2}{\kappa}K_{\mu\nu} \right).
  \qquad
\end{equation}
This is the relation satisfied by the distorted field $\Upsilon$ in \eqref{upsdef}, where both $m^2$ and $p_v^2=\kappa$ are to be understood strictly as constants, with $\omega_m =  \omega - \frac{m^2}{2\kappa} \mu u$. In consequence,  the above-modified  Noether procedure has a Hamiltonian counterpart and the constraint $\phi_3$ is necessary for its realization. Note that  \eqref{modconf} can be reproduced for a generic metric with a (reducible or irreducible) Killing tensor $K_{\mu\nu}$, which means that this type of extended family of conserved charges may emerge in other cases,  not just for pp-wave geodesics.

The intermediate situation where the $p_v^2$ in $Q$ is not considered as a constant $\kappa$ but as dynamical, needs only the constraint $\phi_2\approx 0$ to be satisfied, i.e. $\{Q,\mathcal{H}\} \propto \phi_2 \approx 0$. On the contrary, the integral of the motion obtained by additionally substituting $m^2 = - g^{\alpha\beta}p_\alpha p_\beta$ commutes directly with the Hamiltonian without the need of any constraint. The generic properties of this conserved quantity are studied explicitly in the next section.

To sum up, we have demonstrated the existence of a higher order (non-point) symmetry that yields an integral of the motion which is rational in the momenta. The subsequent use of the constraints $\phi_2\approx 0$ and $\phi_3 \approx 0$ is what helps us reduce the latter to an (on mass shell) equivalent linear expression generated by the distorted conformal vector \eqref{upsdef} that we obtained through our modification. In Table \ref{tab2} we collect the resulting expressions.
\begin{table}
\begin{tabular}{|c|c|c|}
  \hline
  Generator & Conserved charge in phase space & Necessary conditions \\ \hline
  $Y+\frac{m^2}{\kappa} f$ & $Y^\alpha p_\alpha + \frac{m^2}{\kappa} f^\alpha p_\alpha$ & $\phi_2\approx0 \; , \; \phi_3 \approx 0$ \\
  $Y+\frac{N^2 m^2}{\dot{u}^2} f$ & $Y^\alpha p_\alpha + m^2 \frac{f^\alpha p_\alpha}{p_v^2} $ & $\phi_2 \approx 0$ \\
  $Y-\frac{H \dot{u}^2+ 2 \dot{u}\dot{v} + \delta_{ij}\dot{x}^i \dot{x}^j}{\dot{u}^2} f$ & $Y^\alpha p_\alpha - \frac{g^{\mu\nu}f^\alpha p_\alpha p_\mu p_\nu}{p_v^2}$ & - \\
  \hline
\end{tabular}
\caption{The integrals of the motion involving conformal Killing vectors $Y$ for pp-wave geodesics and the conditions needed to commute with $\mathcal{H}$. The constraints allow for a complicated rational integral of the motion (bottom line) to be expressed in linear form (first line).} \label{tab2}
\end{table}

Finally,  let us briefly consider the mass-distorted metric
\begin{equation}
  g^{(m)}_{\alpha\beta}dx^\alpha dx^\beta = dx^2 + dy^2
+ 2 du dv +  \big(H(u,x,y)+m^2 \big) du^2 ,
\label{gmmetric}
\end{equation}
emanating from the right hand side of \eqref{modconf}. A remarkable observation is that the  mass-$m$ geodesics of the metric $g$ in \eqref{e17} are in fact \emph{massless geodesics of the mass-distorted metric} \eqref{gmmetric}.
To see this we fix $m$ and consider the extended Lagrangian \eqref{e8} for a geodesic with mass parameter $M_0$ (to be fixed later) in  the deformed  metric \eqref{gmmetric}
\begin{equation}
  \widetilde{L}_m = \frac{1}{2N} g^{(m)}_{\alpha\beta}\dot{x}^\alpha \dot{x}^\beta - \frac{M_0^2}{2}N \,.
\end{equation}
By inspection, the equations \eqref{xeul}--\eqref{ueul} for the metric $g$ are identical to the corresponding Euler-Lagrange equations of $\widetilde{L}_m$. But we also have the constraint equations
\besub
\begin{align}
\label{constraintLm}
  \left(H(u,x)+m^2 \right) \dot{u}^2+ 2 \dot{u}\dot{v} + \delta_{ij}\dot{x}^i \dot{x}^j + N^2 M_0^2 =0
\qquad  &\text{for}\; g^{(m)}\; \text{with mass}\;M_0
  \\
\label{constraintL}
  H(u,x) \dot{u}^2+ 2 \dot{u}\dot{v} + \delta_{ij}\dot{x}^i \dot{x}^j + N^2 m^2 =0
\qquad    &\text{for}\; g \;\;\;\, \text{with  mass}\;m\,.
\end{align}
\label{twoconstraints}
\esub
The Euler-Lagrange equation for $v$  implies  $N=\frac{\dot{u}}{p_v}$ in both cases. Substituting this into \eqref{constraintL} it becomes identical to \eqref{constraintLm} if $M_0=0$, proving our statement.

\subsection{Conserved charges and canonical symmetries} \label{lcgs}

In this section we discuss the relation between conformal transformations and conservation laws  in the Hamiltonian framework, based on the affine parametrisation. Such an approach has some  advantages: we are dealing with the unconstrained Hamiltonian formalism thus we do not need  to  use the notion of  conditional symmetry. (Ordinary canonical transformations are admitted). Moreover, returning to the  Lagrangian formalism is straightforward.
\par
We  start with the Hamiltonian $\mH_a=\frac12 g^{\mu\nu}p_\mu p_\nu$  on the phase space $(x,p)$  equipped with the Poisson bracket   $\{x^\mu,p_\nu\}=\delta^\mu_\nu$.
Then  given an arbitrary conformal field $Y$ one finds that $G_Y$, defined as
\beq
G_Y= G_Y(x,p)=Y^\mu(x) p_\mu,
\eeq
obeys
\beq
\label{g1}
\{G_Y,G_{Y'}\}=G_{[Y',Y]}, \qquad \{G_Y,\mH_a\}=2\omega_Y\mH_a,
\eeq
and $G_Y$ generate canonical transformations on the phase space. The algebra generated by $G_Y$'s is (anti) isomorphic to the relevant conformal algebra. However, $G_Y$ are conserved if either $Y$ is a genuine Killing field, $\omega_Y=0$, or if we are considering trajectories which lie on the invariant submanifold $\mH_a=0$. In the general  case the second equation  \eqref{g1} allows us to construct the new conserved quantity
\beq
\label{g1a}
\widetilde{G}_Y=G_Y-2\mH_a\int^\sigma\!  \omega_Y(\tilde \sigma)d \tilde \sigma\,.
\eeq
$\widetilde{G}_Y$ is a non-local expression in general. However in  some particular cases, as, for example, for pp-metrics it can be local. In such cases we obtain
\beq
\label{locphase}
\widetilde{G}_Y=G_Y-\Omega_Y\mH_a+\textrm{ irrelevant  terms, }
\eeq
where $\Omega_Y=\Omega_Y(x,p)$ is a function on  the phase space and the ``irrelevant terms" are conserved separately (in other words, to make  the integral in \eqref{g1a} local we can  use the equations of the motion, thus both forms of $\widetilde{G}_Y$  can differ by other  integrals  of the motion).
This local integral of the motion generates a canonical symmetry transformation,
\beq
\tilde \delta x^\mu =\epsilon\{x^\mu,\widetilde{G}_Y\},\quad \tilde \delta p_\mu =\epsilon\{p_\mu,\widetilde{G}_Y\}.
\eeq
One has also
\beq
\label{g2}
\{\widetilde{G}_Y, \widetilde{G}_{ Y_1}\}=\{G_Y, G_{ Y_1}\}+ (\ldots) \mH_a.
\eeq
We conclude that: (i)  on the submanifold $\mH_a=0$  the algebras generated by $G_Y$'s and $\widetilde{G}_Y$'s are isomorphic    and they are (anti) isomorphic to the conformal algebra
(ii)  the canonical transformation on the phase space  generated by $G_Y$'s and   $\widetilde{G}_Y$'s are in general different, even when restricted to the submanifold $\mH=0$; (iii) $\widetilde{G}_Y$ leaves invariant all submanifolds  $\mH_a=-\frac{m^2}{2}$  while  $G_Y$  only leaves invariant the    submanifold $\mH_a=0$  (except when $Y$ is a genuine Killing field).

In view of the above discussion the crucial point is the possibility of localization,  cf. \eqref{locphase}. The more we know about the solution of the equation of the motion the more likely is that we can resolve the localization problem. When the solutions are  known explicitly,  the localization problem can be solved immediately ;  in such a case  all integrals of the motion  are explicitly known  so our procedure  is then  not a very useful one.

However, there exists another possibility: due to the special form of the conformal factor  only  some  partial information  about  the solutions is necessary -- and such information may be available,  because (for example) of other conservation laws. Then we may construct new conservation laws by combining  the conformal transformations with  the already  known conservations laws.  This is what happens  for pp-waves.  Namely, in this case $\Omega$  is simply,
\beq
\label{g3}
\Omega_Y(x,p) =\frac{2}{p_v^2}f^\mu p_\mu ,
\eeq
where $f=f(x,p) $ is defined  by \eqref{derf} with the replacement $m^2\rightarrow -2\mH_a$ (cf. eqn. \eqref{g1a}), and it gives  an ordinary integral of the motion $\widetilde{G}_Y$.

The non-point, canonical symmetries generated by the $\widetilde{G}_Y$'s  can be put  in the Lagrangian form if the momenta are replaced by the appropriate combinations of velocities; however,  then the infinitesimal transformations involve, in general, also velocities (in contrast to   Noether  point symmetries).

To conclude  this  section let us discuss  whether any  new information is carried by the ``conformal" charges. For any genuine Killing vector one obtains an integral of the motion. Thus for  sufficiently symmetric space-time the number of independent integrals of the motion associated with the Killing vectors can  attain  the maximal value  which, for four-dimensional manifolds, equals  7. Then the dynamics governed by $\mH_a$  is superintegrable.

Any additional integral of the motion is a function of those basic ones. Such a situation takes place  for the flat Minkowski space-time~: we have  10 Killing vectors corresponding to the Poincar\'e symmetry. The components of four-momenta $p_\mu$ and the boosts, $M_{0a}, a=1,2,3$ form 7 independent integrals. One can   verify by explicit computations that  the charges associated with all conformal generators  (as obtained in the present paper) are  rationally expressible in terms of them (see sec.  \ref{MinkEx}).

However, in the case of general pp-waves  the situation is different.  It turns out that (see \cite{exactsol}) apart from some special cases,  there is  only one Killing vector $\partial_v$  for  generic pp-waves (giving  only one integral of the motion, $p_v$).  On the other hand,  some classes of pp-waves  admit three proper conformal  fields \cite{b6}.  Moreover,  in sec. \ref{cinvem}  we showed that, even for the Minkowski spacetime,  there are electrodynamic  backgrounds   which break the Poincar\'e symmetry but are  preserved by  conformal fields.  The resulting charges  are not functions  of the Hamiltonian  and $p_v$  only and thus  provide explicit examples  where one obtains new information about the geodesics equation from conformal symmetry.  We  believe that this is the main reason for which the formalism considered in this paper may be really useful.

To conclude, let us note that eqs. \eqref{g1}  and \eqref{g2}   imply  that the Poisson bracket of two localisable charges  gives again a localisable one.   Therefore, starting  from one such charge  and taking its Poisson bracket with the charge generated by a Killing vector (which is thus localisable)  one produces another localisable charge. This process can be continued which implies that the structure of the conformal algebra plays an important role for localization.

\section{Examples}\label{ExamplesSec}

\subsection{A free relativistic particle}\label{MinkEx}

For a free particle in Minkowski  space all Christoffel symbols  vanish and the equations of the motion become
\beq
\label{eomLC}
\ddot{x}^\alpha = \left(\frac{d}{d\tau} \ln (\sqrt{-g_{\mu\nu}\dot{x}^\mu \dot{x}^\nu})\right) \dot{x}^\alpha\,.
\eeq
Equivalently, in terms of the  canonical momenta,
\beq
\label{cmomentaLC}
\frac{d p_{\alpha}}{d\tau} = 0
\;\where\;
p_\alpha \equiv \frac{\partial L}{\partial \dot{x}^\alpha} = \frac{m g_{\alpha\beta}\dot{x}^\beta}{\sqrt{-g_{\mu\nu}\dot{x}^\mu \dot{x}^\nu}}\,.
\eeq
Eqn. \eqref{eomLC}  is integrated as
$
\dot{x}^\alpha = N g^{\alpha\beta} p_{\beta} ,
\,
N = m^{-1} \sqrt{-g_{\mu\nu}\dot{x}^\mu \dot{x}^\nu}\,.
$
Using light-cone coordinates this becomes
$ 
\dot{x}^i = N p_{i}, \, \dot{u} = N p_v, \, \dot{v} = N p_u\,,
$ 
where $p_i, p_u, p_v$ are all constants. Skipping the isometries, we consider 5 proper conformal transformations,
\begin{subequations}
\label{pconfalg}
\begin{align}
Y_{D} &= 2u\partial_u + x^i\partial^i  \qquad  \qquad \qquad \qquad  \qquad \qquad \qquad \quad \;\,
({{\cal{L}}_{Y_{D}}} g)_{\mu\nu} = 2\ g_{\mu\nu},
\label{dilationM}\\
Y_{K} &= u^2\partial_u+ux^i\partial^i-\frac{\bm{x}^2}{2}\partial_v   \quad \qquad  \qquad \qquad \qquad \qquad \;\; ({{\cal{L}}_{Y_{K}}} g)_{\mu\nu} = 2 u \ g_{\mu\nu},
\label{expansionM} \\
Y_{C1} &= \frac{\bm{x}^2}{2}\partial_u -vx^i\partial^i - v^2\partial_v   \   \qquad  \qquad \qquad \qquad \qquad \quad \; ({{\cal{L}}_{Y_{C1}}} g)_{\mu\nu} = - 2 v \ g_{\mu\nu},
\label{C1} \\
Y^i_{C2} &= x^i u\partial_u - \big(\frac{\bm{x}^2}{2} +uv\big)\partial^i + x^i (x^j\partial^j) + x^i v\partial_v \qquad \; \;\; ({{\cal{L}}_{Y^i_{C2}}} g)_{\mu\nu} = 2x^i\ g_{\mu\nu}.
\label{C2i}
\end{align}
\end{subequations}
to which  \eqref{e9} associates the seemingly non-local
conserved charges
\beq
I_j = Y_j^\alpha  p_{\alpha} + m^2\!\int\!\omega_j (x({\tau})) N({\tau}) d \tilde{\tau}.
\eeq
The  $Y^\alpha p_{\alpha}$ terms can be written out explicitly  using \eqref{pconfalg}  and \eqref{cmomentaLC}.
The clue to determine the integral term
is  the property that each conformal factor depends on one coordinate only. The integral can therefore be evaluated and yields simple local expressions,
\benu
\item
For the field  $Y_D$ in \eqref{dilationM}  $\omega = 1$; eliminating  $N$ by
 $\dot{u} = N p_v$  yields~\footnote{Choosing instead $\dot{v} = N p_u$ also allowed 
$
I_{D2} = 2u p_u + x^i p_i + \frac{m^2}{p_u}v + \const
$
These expressions are equivalent because $\dot{u}/p_v=\dot{v}/p_u=N$.},
\beq
I_{D1} = 2u p_u + x^i p_i + \frac{m^2}{p_v}u + \const\,
\eeq

\item
For an expansion $Y_K$ in \eqref{expansionM}  we have $\omega = u$, and we may choose $\dot{u} = N p_v$;  the integration yields
\beq
\cJ_K = m^2\omega\!\int \! N({\tau}) d{\tau} =\frac{m^2}{2 p_v} u^2 + \const
\eeq

\item
Similarly for $Y_{C1}$ in \eqref{C1}  $\omega = -v$ and we may choose $\dot{v} = N p_u$ to get
\beq
\cJ_{C1}=m^2\!\int \!\omega (v({\tau})) N({\tau}) d{\tau} =
-\frac{m^2}{2 p_u} v^2 + \const
\eeq

\item
At last for $Y_{C2}$ in \eqref{C2i}  we get, by choosing $N=\dot{x}^i/p_i$,
\beq
\cJ_{C2}=m^2\! \int^\tau\! \omega (\tilde{\tau}) N(\tilde{\tau}) d \tilde{\tau}
=
\frac{m^2}{2 p_i} {x^{i}}^2 + \const
\eeq
\eenu\vskip-3mm
All these expressions are \emph{local}.
The non-trivial distorted conformal vectors  are the $\Upsilon^{\alpha}=Y^{\alpha}+ \frac{m^2}{p_v^2}f^{\alpha}$  with
\vspace{-2mm}
\begin{align}
  f_D & = u \partial_v
\\[4pt]
  f_K & = \frac{u^2}{2} \partial_v
\\[6pt]
  f_{C1} & = \left[-\frac{1}{4} \frac{m^2}{p_v^2} u^2-u v+\frac{1}{4} \left(x^2+y^2\right) \right] \partial_v -\frac{u}{2} x^i \partial_i
\\[4pt]
  f_{C2}^i & = u x^i \partial_v - \frac{u^2}{2} \partial^i\, .
\end{align}
\label{freef}
Note that the mass appears in $f_{C1}$ only.
When $m\neq0$, the non-Killing  $\Upsilon$ vectors do \emph{not} yield a closed algebra in general~:  for example, $[\Upsilon_K, \Upsilon_{C2}^i ]  = \frac{m^2}{p_v^2} \left[u^2 x^i \partial_v - u^3 \partial^i \right]$.

Moreover, the ``mass distorted'' Lie derivative formula (\ref{modconf}) can be confirmed by means  of the vector field $Y_{C1}$ and its distorted conformal factor $\omega_m = -v - \frac{m^2 u}{2 p_v^2}$\,.

Now, following sec.\ref{lcgs}, we rewrite  the above charges in the form
\eqref{locphase} (see  also \eqref{g3}). Then 
the $\widetilde{G}$'s are ordinary integrals of the motion for the relativistic particle in the affine parametrisation. Since such a  system is superintegrable the charges  $\widetilde{G}$'s  should be expressible  in terms of the basic charges (related to the Poincar\'e symmetry, see the discussion in sec. \ref{lcgs}). In fact, after straightforward but tedious computations one  finds that all conformal generators are rational functions of the 7 independent  integrals $p_\mu$ and the boots $M_{0a}, a=1,2,3$; e.g.  for the conformal generator $K$ we obtain
\beq
\label{gkfree}
\widetilde{G}_K=\frac{1}{2\sqrt 2 (p^0)^2(p^0-p^3)}\left((p^3-p^0)M^{i0}-p^iM^{30}\right)^2
\eeq
All these integrals of the motion generate symmetries  which are  not point transformations. Thus on the Lagrangian level  the infinitesimal transformations contain velocities; e.g.  for   the charge \eqref{gkfree} the corresponding transformation of the configuration  space is
\beq
\delta u=0,\quad  \delta v=\frac{1}{2}(-\bx ^2+\frac{u^2}{u'}{{\bx '}^2}) ,\quad  \delta \bx  =u\bx -\frac{u^2}{u'}\bx '
\eeq
and  it leaves the equation ${x^\mu}''=0$ invariant.

\goodbreak
\subsection{The conformally flat isotropic oscillator}\label{Confflat}

Choosing 
\begin{equation}
\label{osciH}
  H(u,x,y) = x^2+y^2
\end{equation}
in \eqref{e17} yields a conformally flat (not Einstein-vacuum) metric.
In ``Bargmann'' terms the latter describes a two-dimensional isotropic inverted harmonic oscillator ; an attractive oscillator  would be obtained by changing the overall sign of $H$ in \eqref{osciH}. 

As the  properties of this case have been studied  by many authors, here we merely list the principal results.
The metric admits  fifteen conformal Killing vectors~: the well-known seven Killing
vectors $Y_J,\, I=1,...,7$ are completed by eight truly conformal ($\omega_J\neq0$) generators
\begin{subequations}
\label{CKVs}
  \begin{align}
    Y_8 & =  x \partial_x + y \partial_y +2 v \partial_v
 \\[6pt]
    Y_9 & = e^{2u} \partial_u + e^{2 u} x \partial_x  + e^{2 u} y \partial_y -e^{2 u} \left(x^2+y^2\right) \partial_v
 \\
    Y_{10} & = -e^{-2 u} \partial_u + e^{-2 u} x \partial_x  + e^{-2 u} y \partial_y + e^{-2 u} \left(x^2+y^2\right) \partial_v
\\
    Y_{11} & = - (x^2+y^2)\partial_u  +2 v x \partial_x +2 v y \partial_y + \left[2 v^2+\frac{1}{2} \left(x^2+y^2\right)^2\right] \partial_v\\
    Y_{12} & = \; \;\; e^u x \partial_u -\; \frac{e^u}{2}  \left(2 v-x^2+y^2\right) \partial_x \; + \;e^u x y \partial_y \; + \; \frac{e^u x}{2}\left(2 v - x^2- y^2\right)\partial_v
\\
   Y_{13} & =\!-e^{-u} x \partial_u + \frac{e^{-u}}{2}  \left(2 v+x^2-y^2\right) \partial_x + e^{-u} x y \partial_y\! + \!\frac{e^{-u} x}{2}  \left(2 v+x^2+y^2\right) \partial_v
\\
   Y_{14} & = \;\;\; e^u y \partial_u \,+ \, e^u x y \partial_x \; - \; \frac{e^u}{2}  \left(2 v+x^2-y^2\right) \partial_y \,+ \; \frac{e^u y}{2} \left(2 v-x^2-y^2\right) \partial_v
\\
    Y_{15} & =\!-e^{-u} y \partial_u + e^{-u} x y \partial_x +\! \frac{e^{-u}}{2} \left(2 v-x^2+y^2\right) \partial_y +\! \frac{e^{-u} y}{2} \left(2 v+x^2+y^2\right) \partial_v
  \end{align}
\end{subequations}
with conformal factors
\begin{equation}
  \begin{split}
 \label{omegas}
    & \omega_8 = 1, \quad  \omega_9 = e^{2 u}, \quad \omega_{10} = e^{-2 u},
 \\
    &\omega_{11} =2v \quad \omega_{12} = e^u x, \quad  \omega_{13} = e^{-u} x, \quad \omega_{14} = e^u y, \quad \omega_{15} = e^{-u} y \,,
  \end{split}
\end{equation}
yielding the associated integrals of the motion
\begin{subequations}
\label{I8-15}
  \begin{align}
    I_8  = &\frac{m^2}{p_v^2} u+\frac{x \dot{x}}{\dot{u}}+\frac{y \dot{y}}{\dot{u}}+2 v
\\[8pt]
    I_9  = &\frac{1}{2} \frac{m^2}{p_v^2} e^{2 u}+\frac{e^{2 u} \dot{v}}{\dot{u}}+\frac{e^{2 u} x \dot{x}}{\dot{u}}+\frac{e^{2 u} y \dot{y}}{\dot{u}}
\\[8pt]
    I_{10}  = &-\frac{m^2}{2p_v^2} e^{-2 u}-\frac{e^{-2 u} \dot{v}}{\dot{u}}+\frac{e^{-2 u} x \dot{x}}{\dot{u}}+\frac{e^{-2 u} y \dot{y}}{\dot{u}}
\\[6pt]
  \begin{split}
I_{11}  =  &\frac{m^2}{2p_v^2}(\frac{m^2}{p_v^2} u^2 +  4uv -x^2 -y^2) + (\frac{m^2}{p_v^2} ux + 2vx) \frac{\dot{x}}{\dot{u}}+ (\frac{m^2}{p_v^2} uy + 2vy) \frac{\dot{y}}{\dot{u}}  \\
&
- \frac{(x^2 +y^2)\dot{v}}{\dot{u}}+ \big(2v^2 -\frac{1}{2} (x^2 + y^2)^2  \big) \quad \quad
\end{split}
\\[8pt]
\begin{split}
I_{12} = &\frac{m^2 e^u}{4p_v^2}\left( (2u+1)x -(2u-1)\frac{\dot{x}}{\dot{u}} \right) + e^u x \frac{\dot{v}}{\dot{u}} \\
&+ \frac{e^u x}{2} (2v + x^2 +y^2) - \frac{e^u}{2}(2v - x^2 +y^2)\frac{\dot{x}}{\dot{u}}+ e^u xy\frac{\dot{y}}{\dot{u}}
\end{split}
\\[8pt]
\begin{split}
I_{13} =&\frac{m^2 e^{-u}}{4 p_v^2}\left( (2u-1)x +(2u+1)\frac{\dot{x}}{\dot{u}} \right) -  e^{-u} x \frac{\dot{v}}{\dot{u}} \\
&+  \frac{e^{-u} x}{2} (2v - x^2 -y^2)+ \frac{e^{-u}}{2}(2v + x^2 -y^2)\frac{\dot{x}}{\dot{u}}+ e^{-u} xy\frac{\dot{y}}{\dot{u}}
\end{split}
\\[8pt]
\begin{split}
I_{14} = &\frac{m^2 e^u}{4 p_v^2}\left((2u+1)y -(2u-1)\frac{\dot{y}}{\dot{u}} \right) + e^u y \frac{\dot{v}}{\dot{u}}
\\
&+ \frac{e^u y}{2} (2v + x^2 +y^2)
- \frac{e^u}{2}(2v +x^2 -y^2)\frac{\dot{y}}{\dot{u}}+ e^u xy\frac{\dot{x}}{\dot{u}}
\end{split}
\\[8pt]
\begin{split}
I_{15} =&\frac{m^2 e^{-u}}{4p_v^2}\left( (2u-1)y +(2u+1)\frac{\dot{y}}{\dot{u}} \right) -  e^{-u} y \frac{\dot{v}}{\dot{u}} \\
&+ \frac{e^{-u} y}{2} (2v - x^2 -y^2)+ \frac{e^{-u}}{2}(2v - x^2 +y^2)\frac{\dot{y}}{\dot{u}}+ e^{-u} xy\frac{\dot{x}}{\dot{u}}
\end{split}
 \end{align}
\end{subequations}
All integrands are total derivatives  and therefore all quantities are local.

\subsection{A conformally non-flat
vacuum pp wave}\label{ConfNflatEx}

Now we turn to an $u$-dependent vacuum pp-wave metric, \eqref{e17} with
\begin{equation}
\label{funcH}
  H(u,x,y) = \frac{1}{u^4} (x^2-y^2) .
\end{equation}
The regularized version of this metric has been considered before \cite{b3,b4,Conf4GW} ; here we revisit these results using our new framework.

Solving  the conformal Killing  equations $\mathcal{L}_Y g_{\alpha\beta} = 2 \omega g_{\alpha\beta}$ yields five Killing fields ($\omega_i=0$, $i=1,...,5$), we list them for completeness,
\begin{equation}
\label{Kfield}
  \begin{split}
    & Y_1 = \partial_v, \quad Y_2 = \frac{e^{1/u} (u-1) x}{u} \partial_v -e^{1/u} u \partial_x, \quad Y_3 = \frac{e^{-1/u} (u+1) x}{u} \partial_v -e^{-1/u} u \partial_x\,,
 \\[4pt]
    & Y_4 = y \left[\frac{1}{u}\sin \left(\frac{1}{u}\right)+\cos \left(\frac{1}{u}\right)\right] \partial_v -u \cos \left(\frac{1}{u}\right) \partial_y\,,
 \\[6pt]
    & Y_5 = y \left[\sin \left(\frac{1}{u}\right)-\frac{1}{u} \cos \left(\frac{1}{u}\right) \right] \partial_v -u \sin \left(\frac{1}{u}\right) \partial_y\,.
  \end{split}
\end{equation}

The proper conformal fields are the homothety, $Y_6$  and  $Y_7$ in  \eqref{homvfmg} resp.  \eqref{expansionM},
with conformal factors
$  \omega_6 = 1$ and $\omega_7 = u$.
 General theorems say that the maximal number of conformal vectors of a non-conformally-flat pp wave is 7 \cite{exactsol} -- a number which is  attained  in this case.

\begin{figure}[h]
\hskip-7mm
\includegraphics[scale=.3]{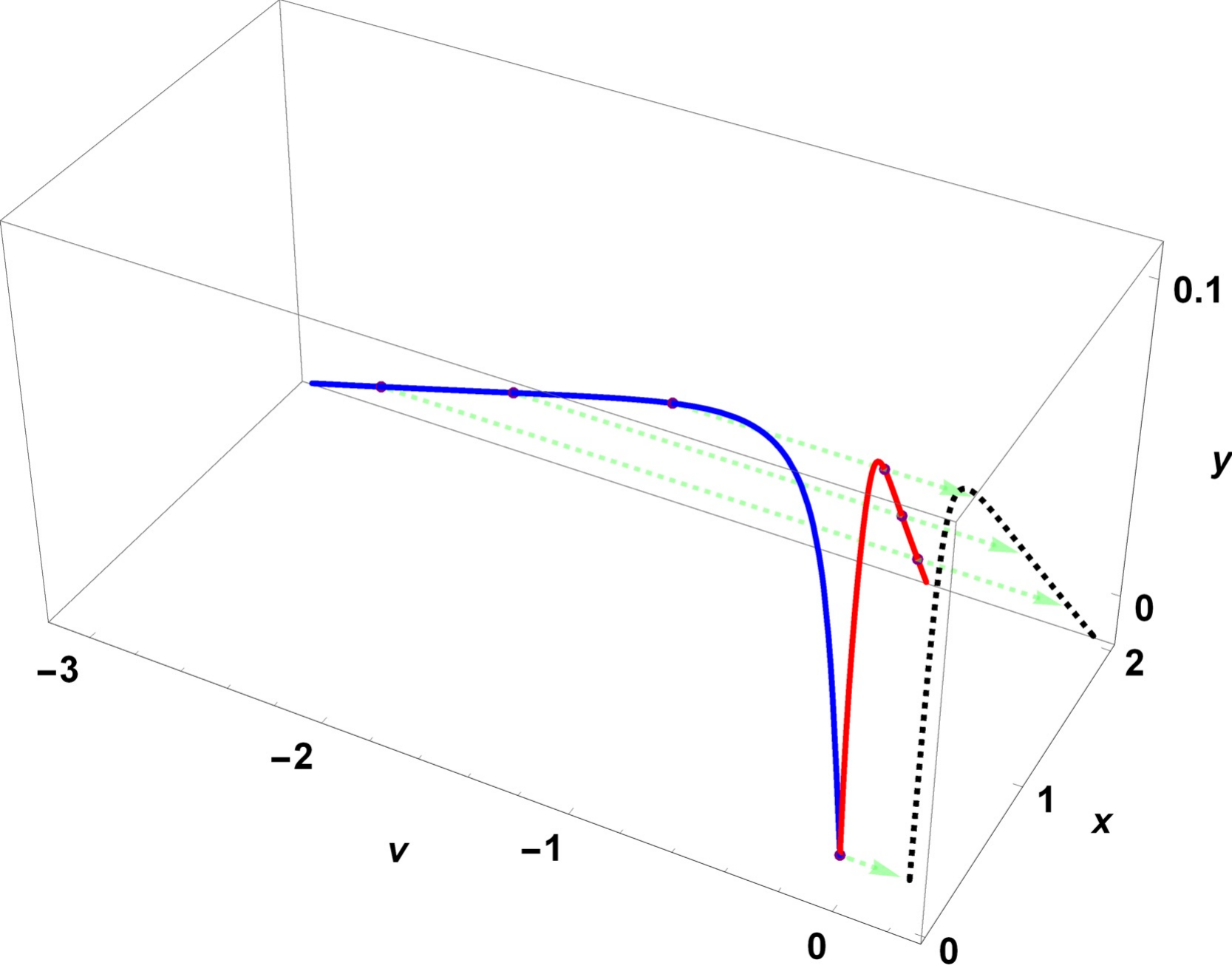}
\null\vskip-3mm
\caption{\textit{\small The massive $m\neq0$ (in \blue{\bf blue}) and massless $m=0$ (in \red{\bf red}) trajectories project on the transverse plane onto the same (dotted) curve. Their $v$-coordinates differ by
$(m/p_v)^2u/2$.
\label{Nonflat}
}}
\end{figure}

The Lagrangian describing a massive relativistic particle moving in such  space-time is given by
\eqref{e8} with Euler-Lagrange equations 
  \eqref{eulgen} and  $H$ defined by  \eqref{funcH}.
The trajectories in the transverse space are plotted in Fig.\ref{Nonflat}.

For the homothety $Y_6$ and the proper conformal Killing vector $Y_7$  eqn. \eqref{e9} with $\phi=0$ yields
the manifestly local conserved charges
\besub
\begin{align}
  I_6 & = \frac{p_v}{\dot{u}} \left( x \dot{x} + y \dot{y} \right) +2p_v v + \frac{m^2}{p_v} u + \text{const.}
\\[4pt]
  I_7 & = \frac{p_v u^2 \dot{v}}{\dot{u}}+ \frac{p_v u \left(x \dot{x}+ y  \dot{y}\right)}{\dot{u}}-\frac{p_v}{2} \left(x^2+y^2\right)-\frac{p_v}{u^2} \left(y^2-x^2\right) + \frac{m^2}{2 p_v} u^2 + \text{const.}
\end{align}
\esub
The mass-dependent terms can actually be eliminated. To see this we observe
that $I_6$ is a combination of $E_N$, $E_i$ (given $N=\frac{\dot{u}}{p_v}$). In particular,
\begin{equation}
  -\frac{d I_6}{d \tau} = 2 \frac{\dot{u}}{p_v} E_N + x E_x + y E_y.
\end{equation}
One can obtain another integral of the motion $I_0$ which does not depend on the mass by  subtracting from $I_6$  (or from $I_7$) the  integrated  equation \eqref{dotvar}  for $\dot{v}$:
\begin{equation}
  I_0 = 2 v + \frac{m^2}{p_v^2} u + \int\!\! \left[\frac{1}{\dot{u}}\left(\dot{x}^2+\dot{y}^2\right) + \frac{\dot{u}}{u^4} \left(x^2-y^2\right)\right] d\tau\,.
\end{equation}
Any linear combination of two integrals of the motion is again  an integral of the motion; in particular we  find
\begin{equation}
  \tilde{I}_6 = \frac{1}{p_v}I_6 - I_0 = \frac{1}{\dot{u}} \left(x \dot{x}+ y\dot{y} \right) - \int\!\! \left[\frac{1}{\dot{u}}\left(\dot{x}^2+\dot{y}^2\right) + \frac{\dot{u}}{u^4} \left(x^2-y^2\right)\right] d\tau .
\end{equation}
$\tilde{I}_6$ does not depend on the mass and it is now - since we eliminated $E_N$ - an integral for  the $E_i$ equations,
$
  \frac{d \tilde{I}_6}{d \tau} = y E_y - x E_x .
$ 

Following the same procedure applied to  $I_7$ we end up with an expression which again  does not depend on the mass~: we get an integral of the motion for  the $(x,y)$ equations alone, which is \emph{exactly the same as in the massless case}. The mass appears in the constraint equation and through it affects only the $v$ variable. Integrals of the motion which do not contain the $v$ variable are independent of the mass.

Finally, let us see what one gets  from the modified  Noether procedure. The symmetry generators \textcolor{magenta}{(\ref{upsdef})} are\vspace{-5mm}
\begin{equation}
\label{symex2}
  \begin{split}
    & \Upsilon_i = Y_i, \quad i=1,...,5 \\
    & \Upsilon_6 = Y_6 + \frac{m^2}{p_v^2} u\,\partial_v \\
    & \Upsilon_7 = Y_7 + \frac{m^2}{2p_v^2} u^2 \partial_v .
  \end{split}
\end{equation}
The Killing vector fields $Y_i$, $i=1,...,5$, yield the  Noether symmetries, while the modification provides us with additional mass-dependent ``corrections" for the conformal Killing vectors.
Obviously $Q_i=I_i$ for $i=1,...,5$ and for the last two $Q$'s \textcolor{magenta}{(\ref{netint})} we obtain
\besub
\begin{align}
  Q_6 & = \frac{1}{N} \left( x \dot{x} + y \dot{y} \right) +2 v \frac{\dot{u}}{N} + \frac{m^2}{p_v^2} u \frac{\dot{u}}{N} &= I_6 \,,
\\[8pt]
  Q_7 & = \frac{u^2 \dot{v}}{N}+ \frac{u \left(x \dot{x}+ y  \dot{y}\right)}{N}-\frac{ \left(\left(u^2-2\right) x^2+\left(u^2+2\right) y^2\right)\dot{u} }{2 N u^2} + \frac{m^2}{p_v^2} u^2 \frac{\dot{u}}{2N}&= I_7\, .
\end{align}
\label{symex2Qchar}
\esub
Lastly we note that the ``distorted'' Killing vectors (\ref{symex2}) and the associated conserved charges (\ref{symex2Qchar}) satisfy (\ref{modconf}) and (\ref{condsym2}), respectively.

\section{Conclusion}
In this work we investigated the  conserved charges  associated with conformal  Killing fields in curved space-times possibly equipped also with an electromagnetic background preserved by those Killing fields. We put  special emphasis  on massive particles -- those which are mostly considered in the Memory Effect \cite{OurMemory}. The associated conserved quantity in \eqref{e4} involves an integral term, eqn. \eqref{intterm}, which requires integration along the trajectory and  could  therefore  be non-local. It is only for a special  parametrisation that this term become local general.  However, such conceptual and calculational difficulties are absent in pp-wave space-times~: the integral term  can be calculated analytically and  becomes local in an arbitrary parametrisation, as implied by \eqref{fe}.

pp-wave space-times play a role for the Eisenhart-Duval lift \cite{Eisenhart,Bargmann,DGH91,dissip} of  2-dimensional classical dynamics. Analysing  the meaning of the charges in this context, we have shown that  after expressing them in terms of transversal coordinates, the term with $m$ reduces to a  constant. This is consistent with the observation that both massive and massless geodesic motion are lifts of the same underlying classical dynamics.  

Moreover, considering conformally related (and consequently physically inequivalent) pp-wave metrics,  we have shown that the conformal charges for massive geodesics coincide.  
Explicit examples  allow us   to give more clear interpretation of some  charges corresponding to  proper conformal fields.
\par Next, we constructed  a family of pp-waves (which include the Minkowski space-time)  and  an  independent  family  of electromagnetic backgrounds which are preserved by  a suitable  conformal field. The  explicit form of the corresponding  charges  was given. We gave  the example of electromagnetic backgrounds for which  the conformal symmetry yields a new integral of the motion.
\par In the usual approach the Killing vectors can be identified  with  Noether point symmetries (and consequently give Noether charges which are linear in the momenta); however, for massive particles, the  conformal  vectors  do  \emph{not}  define symmetries  \cite{Hussain2} in general.

In eqn \eqref{upsdef} of sec.\ref{ModConfpp} we introduced  ``distorted'', non-point transformations, which are analogous to dynamical symmetries and related them to local conformal charges.  First, we rewrote the local charges  in an ``almost  Noetherian'' form (using the parametrisation invariant approach); although the distorted field  contains momenta  and isn't  formally a  point symmetry, fixing $p_v=$const. allows us to interpret it as such.  

Moreover,  in the context of the charges we obtained, we modified appropriately the Noether procedure  by fixing the momentum by a supplementary condition); and discussed its geometric  meaning.
\par
Next,  we analysed  the charges as associated with the symmetries of the canonical Hamilton equations (using the Hamiltonian approach  with affine parametrisation). In  view of these considerations  the conformal  Killing fields,  together with an appropriate distortion, generate  non-point symmetry transformations  and  induce velocity-dependent  transformations in  the  configuration  space.  

At last we discussed the possibilities of localization of  conformal charges and their relevance in integrability of geodesics equations  and  presented  some further  examples.
\par
The process that we followed can be extended in other configurations apart from pp-waves. The results we obtained fit into various  recent studies of the relations between  conformal symmetries  and integrability  problems, sheding  new light at these issues.

\begin{acknowledgments}
 We would  like to acknowledge Xavier~Bekaert and Gary Gibbons for discussions.
 ME thanks the \emph{Denis Poisson Institute of Orl\'eans-Tours University} for hospitality.
This work was partially supported by the the National Natural Science Foundation of China (Grant No. 11975320) and
   National  Science  Centre of  Poland (Grant No.   2016/23/B/ST2/00727).
\end{acknowledgments}
\goodbreak

\end{document}